\documentclass[10pt]{iopart}
\begin{document}

%\title[Numerical Relativity: A review]{Numerical Relativity: A review}

\topical{Numerical Relativity: A review}

\author{
Luis Lehner
}

\address{
Department of Physics and Astronomy  \\
\& Pacific Institute for the Mathematical
Sciences\\
The University of British Columbia, Vancouver, BC V6T 1Z1}

\ead{luisl@physics.ubc.ca}

\date{\today}

\begin{abstract}
Computer simulations are enabling researchers to investigate
systems which are extremely difficult to handle analytically.
In the particular case of General Relativity, numerical models
have proved extremely valuable for investigations of strong field
scenarios and been crucial to reveal unexpected phenomena.
Considerable efforts are being spent to simulate astrophysically
relevant simulations, understand different aspects of the theory
and even provide insights in the search for a quantum theory of gravity.
In the present article I review the present status of the field 
of Numerical Relativity, describe the techniques most commonly
used and discuss open problems and (some) future prospects.
\end{abstract}

\submitto{\CQG}


\maketitle

%\pacs{04.25.Dm, 04.30.Db}
%
%\vskip2pc]
%
%\narrowtext
%

%%%%%%%%%%%%%%%%%%%%%%%%
%%%   INTRODUCTION   %%%
%%%%%%%%%%%%%%%%%%%%%%%%

\section{Introduction}
\label{sec:Introduction}
The beginnings of the twentieth century witnessed a major
revolution in our understanding of gravitation. Einstein's theory radically
changed the way we conceived
gravity and its effects. Unraveling the messages that his theory contains
requires the ability of solving a coupled nonlinear system of ten
partial differential equations. These are `special' equations as they
govern the very structure of the spacetime itself (as opposed to other theories
where the fields evolve `on top' of an unchanging spacetime).

For about six decades, only in special situations were researchers able to 
obtain solutions to these equations. These assumed the existence
of symmetries and/or concentrated on asymptotic regimes that allowed considerable
simplifications of the equations reducing them to a manageable (and solvable)
system. Although certainly considerable `new' physics has been learned from Einstein's
theory, its full implications remain elusive. 

The last decades of the twentieth century witnessed another 
revolution. This one, the `computer revolution', was spurred by the computational
capabilities that powerful computers provided researchers. This
new tool allows the study of systems which would otherwise be impossible (or extremely
involved) analytically. Simulations not only are letting researchers tackle
difficult problems but also allow for a nice visualization of the outcome. These
simulations serve as theoretical laboratories
for General Relativity, where, the past impossibility of
constructing a gravitational laboratory prevented data-driven research
from aiding in our explorations
of the theory. We have already witnessed some of the benefits that these `numerical
laboratories' can provide, for instance, they 
have demonstrated the existence of critical behavior in General Relativity and naked singularities in
gravitational collapse; the possible appearance of toroidal event horizons; indicated generic properties of
singularities in cosmological contexts; provided more accurate understanding of rapidly rotating neutron
stars and shed light into the structure of singularities.

The continuous improvements in computer power coupled with the gained (and being
gained) experience in simulating Einstein equations signal that, after almost
a century, we are on the path of unveiling what these equations have so far kept hidden.

Computer simulations are and will increasingly be of crucial importance to let us
study strongly gravitating systems like those containing massive stars
and/or black holes; spacetimes on the verge of black hole formation; 
investigation of cosmological scenarios, studies of structures of singularities and
even for investigations of different aspects of possible quantum theories of gravity.

Clearly, understanding these issues is very important academically since it will
definitively advance our scientific knowledge. Additionally, a thorough understanding of some of these
systems is also relevant from a ``more practical'' point of view. Technology is also setting us
at the verge of being capable, for the first time, to directly test General Relativity in the strong
field limit and use it to obtain a new window with which to scrutinize our universe.
The beginnings of the twenty first century will witness the operation of several highly sensitive
gravitational wave earth-(and probably space)-based detectors\cite{LIGO,VIRGO,GEO,TAMA,LISA}. These detectors will allow
researchers to study signals produced from strong field systems and therefore will provide
a chance to test a theory which has so far proven very successful in weak regime scenarios. These 
signals carry specific signatures of the system that produced them
and therefore their detection and analysis will represent a new form of astronomy, ``gravitational
wave astronomy''\cite{thornewaves,schutz}. This astronomy will require accurate models of the sources and
the waveforms they produce to decode the information carried out by gravitational
waves. These models will be provided by numerical relativity.

The importance of numerical models of relativistic systems can not be overestimated.
This has been reflected in the growing interest in numerical relativity since its
first tentative steps in the late 60's. Perhaps this growth is better described
by noting that
a review on the status of numerical relativity 30 years ago would have been dedicated to describe
what the pioneers of this field,
Hahn \& Lindquist\cite{hahnlindquist64}; Smarr\cite{smarr78} and  Eppley
\cite{eppley77} studying black hole spacetimes and Wilson\cite{wilson} investigating neutron stars
were doing back then.  These pioneers foresaw the 
importance of
computers in modeling otherwise intractable problems. These first ventures
spurred throughout the years a large number of projects in many different directions.
As a consequence, a comprehensive review would demand a complete edition 
of Classical and Quantum Gravity to justly describe most efforts and
directions being studied at
present. This, naturally, speaks well for the status of the field, signaling
how much momentum has gathered in the past decades and how an increasingly important role
is playing in today's gravitational research. Unfortunately, lack of space will
not allow for fairly addressing all `flavors' of numerical relativity research. 
The vast number of areas renders covering all them impossible; as much as I tried presenting
a comprehensive overview, some topics or a more detailed presentation of others 
are not included and I apologize in advance for this.
In particular, I very much regret
not being able to extensively cover areas like
Relativistic Hydrodynamics, Critical Phenomena and Computational Cosmology in this article. Fortunately,
excellent recent reviews are available on these subjects (and I will refer the reader
to them as I briefly go through the subjects). This review should be considered 
complementary to these. I will put more emphasis on areas which I consider 
basic to understand the present status of the field (and that are common to all areas of
numerical relativity) and to serve as a guidance to researchers
and students willing to immerse in this wonderful (relatively) new discipline in G.R. \\

The {\it main goal of numerical relativity} is to provide the description of the spacetime by 
solving Einstein equations numerically. This numerical implementation provides
the metric $g_{ab}$ on, at least, some region
of the manifold $\cal M$  (${\cal M}$ being an orientable, n-dimensional manifold of all
physical events and $g_{ab}$ a Lorentzian metric tensor).
This manifold is assumed to be simply connected and globally hyperbolic, therefore, given
appropriate data on an initial hypersurface, its future development can be obtained
by means of solving Einstein equations\cite{wald}. [Although analytical extensions of
non-globally hyperbolic formulations can be obtained, the numerical treatment of such situations
is much more complex and has so far not been considered].

Perhaps an obvious point sometimes overlooked when thinking of numerical models
to solve a given problem is that {\it computers are not magic!} Although our computational
resources give us a powerful tool with which to attempt solving a
problem, it certainly does not provide a magical solution. One must worry
about the `standard points' proper of the traditional  `pencil and paper method'
but also keep in mind that a numerical simulation will be employed, which
adds a new dimension to the specification of the problem.
Hence, before attempting any computation one must carefully 
\begin{itemize}
\item {Choose appropriate form of equations and set of variables that govern the system}
\item {Adopt a suitable reference frame with respect to which describe the system}
\item {Define initial and/or boundary conditions }
\end{itemize}
In a numerical approach, the aforementioned points
should be chosen in a way that will possibly aid, or at least not harm, 
the numerical implementation. This introduces a new set of choices
\begin{itemize}
\item {Discretization strategy}
\item {Specific Algorithms}
\end{itemize}
I will organize the presentation following this rather natural path. I review in section II
the basic arena, giving an introductory description about the issues involved in obtaining
the system of equations, coordinates choice and initial and boundary conditions. Then, in
section III, a more detailed presentation of the three main avenues towards implementing
Einstein equations presently employed is presented. In each case, a particular representative system
is discussed as an example, how coordinate systems can be chosen and the initial and boundary
values specification is addressed (Here for the sake of clarity I will concentrate on the vacuum case). 
Section IV is devoted to some generic aspects related to numerical
techniques while section V to particular issues related to the numerical implementations
(separately addressing particulars of the three avenues presented in section III). In section VI, I 
discuss the main aspects and
techniques related to non-vacuum problems. Then, in section VII a (partial) list of the main
past accomplishments of the field are presented while section VIII comments on the major
current problems and results. Towards the end, in section IX, I describe a few efforts towards
employing numerical simulations as a complementary technique to fully describe binary systems
from their very early stages to the final merged object. Finally, in section X, I briefly
comment on the main problems for the future and conclude in section XI. \\

Note: When writing this article I had three audiences in mind. Researchers
outside the field who
just want to get a current glimpse on the main issues and approaches of the field to whom
I would recommend sections I through III, VII, VIII, X and XI. Another one of those interested in 
getting involved in Numerical Relativity, who additionally might find sections IV and VI useful
in `breaking the ice'. And finally practitioners of the field who I hope will benefit from a comprehensive literature
survey throughout the article, specific discussions in sections V and IX and the `broad picture'
of future possible directions presented in section X.\\

Throughout this paper I adopt
geometric units where $G=c=1$. Additionally, small case Latin letters in the first half of the
alphabet range from 1 to 4 and those from $i$ on range from 1 to 3, unless
otherwise indicated.

\section{The arena}
{\bf System of equations}\\
The theory of General Relativity clearly stands out from all others by the fact
that the spacetime, defined as the pair $({\cal M},g_{ab})$  is `obtained' from Einstein equations
all at once. What one solves for is the geometry, not for a particular metric tensor
(since two tensors differing by a diffeomorphism describe exactly the same geometry). The 
`unknown variables' do not `live' on top of the spacetime, but rather they are precisely the spacetime.
Hence, right from the start, the problem of even posing the equations 
is not a straightforward one. Einstein equations, $G_{ab}=8 \pi T_{ab}$, 
(with $G_{ab}$ the Einstein tensor and $T_{ab}$ the stress energy tensor) are completely independent
of any coordinate system. The lack of a preferred frame of reference, which is a natural manifestation of
the equivalence principle, is at the very core of the theory. The complete freedom
in the frame choice is in practice exploited to express the equations in a more convenient
way which has lead to several formulations of General Relativity. Roughly speaking, a notion of
time is introduced and the level surfaces defined by this time can be spacelike (giving rise to
a {\it `3+1' or Cauchy approach to GR}), null (defining a {\it characteristic approach}), or of more generic
type (which yield the approaches like the {\it conformal Einstein equations}; 
{\it Cauchy-characteristic matching}, etc).\\

Once the system of equations is chosen, as is the case with any simulation, care must be taken 
with adopting (I) a preferred set of suitable coordinates (so that from the equivalence class of
metric tensors defining the same geometry a single one is obtained) 
and (II) appropriate initial and boundary data for the problem under consideration.\\

{\bf Suitable Coordinates}\\
When Einstein equations are recast in a way amenable to a dynamical description, one coordinate, say $x^0$,
is chosen to play the role of `time' with respect to which the dynamical evolution will referred to.
Then, n-1 additional coordinates, $x^i$ ($i=1..n-1$), are introduced at the level surfaces ($\Sigma_t$) of the
time parameter. These coordinates could be standard like Cartesian; spherical, cylindrical, etc. or others better
suited for specific problems. Note that one still has quite some freedom left, namely the rate of change of the
time coordinate need not be uniform as a function of $x^i$. Additionally, the $x^i$ at different values
of the time coordinates might not be constant along the direction normal $\Sigma_t$, ie. might be ``shifted''. 
Exploiting this freedom has proven useful in numerous analytical studies (eg. the use of harmonic coordinates
render Einstein equations in an explicitly hyperbolic form which is convenient to
analyse properties of the expected solution). In numerical implementations this freedom can prove crucial
and the adoption of convenient coordinates is a very delicate (and important) problem which
has no `clear cut' solution. These ``ideal''  coordinates satisfy the following properties

\begin{itemize}
\item{{\bf Singularity avoidance properties (A) or amenability for singularity excision (B):} 
Spacetimes containing singularities can be approached by either ``slowing down''
the rate of time change in a region near the singularity so that the evolution is ``frozen'',
thus avoiding the evaluation at singular regions (A) or excising the singularities
from the computational domain, thus getting rid of the problematic region, this can safely be done
assuming the singularities are not ``naked'' due to the event horizon hiding the excision process (B).}
\item{{\bf Simplification of variables:} Properly chosen coordinates might simplify the metric tensor. For instance,
in the presence of a symmetry, by choosing a coordinate adapted to the congruence defined by such symmetry
the metric tensor does not explicitly depend on such coordinate.}
\item{{\bf Degrees of Freedom:} Adopting coordinates that manifest the true degrees of
freedom might help obtaining accurate physical predictions.}
\item{{\bf Radiation Propagation:} When gravitational waves are sought for, coordinates
adapted to a natural radiation gauge can
 considerably simplify the numerical treatment\cite{bardeenHOUCHES,bondi,smarr78}}
\end{itemize}
With prior knowledge of the dynamics of the system it is certainly easy to come up with coordinate 
prescriptions satisfying these properties. However, we need numerical simulations
to obtain this knowledge! A great deal of effort has been put into obtaining reasonable recipes to appropriately
choose coordinates and I will outline several proposals in this direction. However, our present knowledge
on this subject is still rather limited, the field would certainly benefit from further research in this direction.\\

{\bf Initial Conditions}\\
Specification of the initial and boundary data determine the physical situation
under study. In General Relativity, a theory with only two degrees of freedom `hidden' in
the six components of $g_{ab}$ (assuming four are fixed by coordinate conditions),
it is not expected that all can be specified freely at the initial time; rather, there must be
constraints limiting the possible choices. Consequently, before starting the evolution problem, 
one must take care of the initial value specification which requires careful examination of the
constraint problem. Additionally, even when the equations defining consistent initial data can
be readily solved (in terms of some freely chosen functions), these must be chosen so that they
represent the targeted physical system.\\

{\bf Boundary Conditions}\\
As important as the initial value specification is that of the treatment of
the possible boundaries. These boundaries can be at the `outer edge' of the
computational domain (referred to as {\it outer boundaries}) or inside the 
computational domain (referred to as {\it inner boundaries}).
Not only must the prescription of boundary data correspond
to the physical situation in mind, but also its implementation not give rise to spurious reflections
which could contaminate the described physics or, even worse, render the simulation unstable.
Properly addressing the boundary implementation is a highly non-trivial problem even in simple systems. For
instance, when modeling the simple wave equation in dimensions higher than one, correct boundary value 
specification requires a non-local procedure which represents a significant computational
overhead\cite{waveHIGHDBOUND}.
In nonlinear systems, where backscattering is expected, this problem becomes very difficult and a general solution
is not known even at the analytical level. Clearly, the numerical treatment of the boundary value problem
is a delicate issue, and I will review the present way of handling it in the next sections.
\\
\\
 
In the following section I will comment on how the above mentioned problems are addressed in the different
formulations that have made their way into Numerical Relativity\footnote{For the sake of keeping the presentation
short, I will restrict to the vacuum case until section~\ref{matter}. However, most of what I describe
here applies to the  non-vacuum case, the additional problem is the accurate treatment of the
equations governing the matter variables.}.

\section{Formalisms; initial/boundary data and coordinate conditions} \label{formalisms}

\subsection{Cauchy approach to GR \label{form:cauchy}} 

\subsubsection{Formalism:}
In the 3D Cauchy (or ``3+1'') formulation of Einstein's equations, one
foliates $\cal M$ with a parametrized (with parameter $t$) set of spacelike,
3-dimensional hypersurfaces $\Sigma_t$ and chooses coordinates $x^i$ (i=1..3)
 to label
points on each one. Thus, the spacetime points have coordinates $x^a=(t,x^i)$.
The standard 3+1 decomposition presented
in~\cite{mtw,york-sources,choquet-york-GRG}, chooses $n^\mu$
as the future-pointing timelike unit normal to the slice, with
\begin{equation}\label{eq:unit-normal}
	n^\mu \equiv -\alpha\nabla^\mu t \, ,
\end{equation}
$\alpha$ is the \emph{lapse function} defining the proper interval measured
by observers traveling normal to the hypersurface. Since coordinates need not be
chosen to remain constant along the normal direction (as they can freely specified at
each $\Sigma_t$), they are related by a {\it shift vector} defined as
\begin{equation}\label{eq:time}
	\beta^\mu \equiv t^\mu - \alpha n^\mu,
\end{equation}
where
\begin{equation}\label{eq:shift-const}
	\beta^\mu n_\mu \equiv 0 \, ;
\end{equation}
so, in this frame, $\beta^a=(0,\beta^i)$.
If the (Euclidean) metric of each $\Sigma_t$ is given by $\gamma_{ij}$ (defined
as the pull-back of $g_{ab}$ onto  $\Sigma_t$)
the spacetime metric results
\begin{equation}\label{eq:3+1-interval}
	{\rm d}s^2 = -\alpha^2{\rm d}t^2
		+ \gamma_{ij}({\rm d}x^i + \beta^i{\rm d}t)
				({\rm d}x^j + \beta^j{\rm d}t).
\end{equation}
$\gamma_{ij}$ is regarded as a fundamental variable while $\alpha$
and $\beta^i$ mere manifestations of the coordinate freedom proper of General Relativity.
When writing down Einstein equations in this approach, a second order PDE system
results where, in particular, six equations contain second time derivatives 
of $\gamma_{ij}$ (obtained from $G_{ij}=8\pi T_{ij}$). In order to properly specify the
 initial value problem, the first time
derivative of $\gamma_{ij}$ must be also specified at an initial hypersurface.  Instead of
this, one usually provides $K_{ij}$ defined by
\begin{equation}\label{eq:extr-curv}
	K_{ij} \equiv -{\textstyle\frac12}{\cal L}_n\gamma_{ij},
\end{equation}
where ${\cal L}_n$ denotes the Lie derivative along the $n^\mu$ 
direction. From $K_{ij}$
the first time derivative of $\gamma_{ij}$ is readily obtained but $K_{ij}$ is preferred (as it has
a natural geometrical interpretation, being the second fundamental form or extrinsic curvature of the
$\Sigma_t$ embedded in the four-dimensional spacetime).
With these definitions, Einstein equations are expressed (with the aid of the Gauss--Codazzi--Ricci conditions)
as
\begin{eqnarray}
	d_t \gamma_{ij} = -2\alpha K_{ij} \, ; \label{3plus1h} \\
	d_t K_{ij} = \alpha \left[
		{R}_{ij} - 2K_{i\ell}K^\ell_j + K K_{ij}\right] - D_i D_j\alpha \, ; \label{3plus1k} 
\end{eqnarray}
where, $d_t \equiv \partial_t - {\cal L}_{\beta}$; $D_i$ and ${R}_{ij}$ are the 
covariant derivative and Ricci tensor compatible with $\gamma_{ij}$ and 
$K\equiv K^i_i$.  

%%%%%%%%%
Hence, $\gamma_{ij}$ and $K_{ij}$ are the  set of initial
data that must be specified for a Cauchy evolution of Einstein's
equations.
Equations~(\ref{3plus1h}) and (\ref{3plus1k}) constitute
the evolution equations which are used to obtain the spacetime to the future
of the initial hypersurface. There still remains four extra equations
which we have so far not considered (from $G_{0i}=8\pi T_{0i}$, which do not contain second time
derivatives of $\gamma_{ij}$). These equations are
\begin{equation}
	{R} + K^2 - K_{ij}K^{ij} = 0 \, ,\label{eq:Hamiltonian-const}
\end{equation}
and
\begin{equation}
	D_j\left(K^{ij} - \gamma^{ij}K\right) = 0. \label{eq:momentum-const}
\end{equation}
Equation~(\ref{eq:Hamiltonian-const}) is referred to as the
{\it Hamiltonian} or {\it scalar constraint}, while (\ref{eq:momentum-const}) are
referred to as the {\it momentum} or {\it vector constraints}.  These equations 
impose conditions that $\gamma_{ij}$ and $K_{ij}$ must satisfy and therefore
restrict their possible values. Fortunately, only at the initial hypersurface
must one worry about satisfying the constraint equations as the Bianchi identities
guarantee they will be preserved on future slices of the evolution. Providing data
satisfying the constraint equations is not a trivial task, we will return to this issue in
section~\ref{initial_data}.

This Cauchy or `3+1' formulation is customarily called ADM in numerical relativity jargon\footnote{For
the formulation introduced by Arnowitz, Deser and Misner\cite{mtw}; although
it is related to it by using $K_{ij}$, instead of the ADM conjugate momentum $\pi_{ij}$.} and has been the system
until recently has received the most attention in Numerical Relativity. However, 
this system is by no means the only `3+1' approach. Many related formulations 
can be readily obtained from the ADM. For instance, one can choose (i) to use a different combination
of variables;
(ii) the constraints can be freely added to the equations (premultiplied by arbitrary
functions) and (iii) extra variables can be introduced to eliminate second order spatial derivatives (with
the consequent enlargement of the system of equations). [Note that these in turn can be expressed in terms
of tensor, frame or tetrad components].

Several of these options have been exploited to come up with new, and of course, physically 
equivalent re-formulations which explicitly display some desirable properties. Among those, a number of 
symmetric hyperbolic formulations\footnote{For a recent review of hyperbolic systems in General Relativity 
see~\cite{reulaREVIEW}.} have been presented (using (i)-(iii)) and are starting to make their way into Numerical
Relativity (see for instance \cite{spectral2,arbonaGAUGE,shinkaiASHTEKAR,bardeenPRIVATE,generalEC}). These formulations are
written in first order form and the standard mathematical machinery
for PDE's can be used to determine the well posedness of the problem under study;
whether the characteristic speeds of the system are physical (lie inside the null cones) and furthermore, determine which
combination of variables are ingoing and outgoing with respect to a given boundary. This
 plays an important role when imposing boundary
conditions (see next section). Additionally, other, `less ambitious' systems [obtained using (i)-(ii)] aimed 
towards isolating the physical modes of the solution have recently become quite popular
in Numerical Relativity. This approach known as BSSN is displaying in a number of
cases better behaved evolutions than those obtained with the ADM 
formulations\cite{shibnakam,baumgarteBOUND,golmEVOL}\footnote{Systems of this type have
 also been introduced which can be rendered
symmetrically hyperbolic by appropriately adding the constraints\cite{frittelliCADM}.}.\\

\subsubsection{Coordinate conditions:}
In this approach, adopting coordinates conditions means providing a prescription
for $\alpha$ and $\beta^i$ (the lapse and shift vector). One would like this prescription to be ideally suited
for the simulation; however, as mentioned
previously, this is not generally possible without prior knowledge of the expected dynamics. To achieve this,
one can somehow `tie' the coordinate conditions to the dynamics of the fields so as
to obtain some `feedback' on how these coordinates should be chosen.
In practice, either `evolution' equations or  equations at a given surface (elliptic) are employed for this purpose.
The former approach, although sound in principle, should be treated with special care,
as some choices might lead to coordinate pathologies \cite{bruhatGAUGE,golmCOORDS,hernGAUGE}.
The latter option involves solving elliptic equations which are computationally expensive,
nevertheless have proven quite useful. In the following we will present some of the
options being pursued.  

These can be grouped into three different strategies: ``geometrical''; ``simplifying'' and
``cost-reducing" conditions. The division between the first two
is clear in methodology but not necessarily in the final results since, as we will see next, 
some conditions are obtained with either strategy. In the third group, I gather computationally less
expensive conditions defined (I) to retain some of the properties of those in the first two groups while 
at the same time simplify their numerical implementations or (II) derived from known
solutions. \\

{\bf``Geometrical'' prescriptions}\\

{\it Lapse condition: Maximal slicings} \\
The first of these prescriptions was suggested by Lichnerowicz\cite{lichnerowiczID} and
later extended by York\cite{york-sources},
known as the family of `maximal slicings'. These slices maximize the 3-volume of the
slices, hence the name. This condition translates into slices that 
effectively deform so 
that $K \equiv \gamma^{ij} K_{ij}\equiv F(t)$ which in turn implies 
a non-uniform $\alpha$. A straightforward evaluation of the trace of equation (\ref{3plus1k})
(and using the Hamiltonian constraint to re-express the Ricci tensor in terms of
$K_{ij}$), provides the elliptic equation for $\alpha$,
\begin{equation}
\Delta \alpha = \alpha  K_{ij} K^{ij} - K_{,t} \, .\label{maximalgauge}
\end{equation}
Although it is not clear that a solution to the above equation will always exist, in present
and past applications (in the particular case of $K=0$) it has proven quite useful. Not only does
it provide a usable definition for the lapse, but the resulting slicing
tends to `avoid the singularities'\cite{york-sources}. Note that from equation (\ref{3plus1h}) one straightforwardly
obtains the equation $\partial_t (log \sqrt{\gamma})=-\alpha K + D_i \beta^i$ which describes
the evolution of the determinant of $\gamma_{ij}$. In the case where $\beta^i=0=K$ the singularity avoidance
property of this slicing is clear as the variation of the local volume remains fixed.
This effectively
slows down the evolution in regions of strong curvature while the simulation proceeds in the farther regions.
Unfortunately, this feature comes at a price. The same property that makes it enticing
carries the crux when attempting long simulations of singularity-containing spacetimes. As the evolution
 proceeds, the slices
``pile-up'' in regions of  high curvature while not in weaker curvature regions. The sequence of slices 
result considerably `bent' and 
large numerical gradients are induced (this problem is usually referred to as 'grid stretching'; however the grid
clearly does not stretch, rather the proper distance between grid points become large). As the 
evolution proceeds these gradients become larger and ultimately
the evolution crashes. In almost all implementations employing maximal slicings, the choice of $K=0$ has been
adopted. Recently, the properties of slices with non-vanishing $K$ have been analyzed in 1D
illustrating the potential advantages of such choice\cite{holzCRUNCH}. \\

{\it Shift conditions: Minimal Strain and Minimal Distortion} \\
A shift condition known as `minimal strain' was introduced by Smarr and York\cite{smarryork} through a
set of elliptic equations obtained via a minimization of the hypersurface strain. Minimizing an
action defined with $g_{ij}$ and ${\cal L}_{n} g_{ij}$ with respect to  $\beta^i$ 
yields the (elliptic) set of equations,
\begin{equation}
D_i D^i \beta^j + D_i D^j \beta^i - 2 D_i(\alpha K^{ij}) = 0\, . \label{ibbhSHIFT}
\end{equation}
A related condition known as `minimal distortion' is obtained by considering 
a different action defined in terms  
of a ``distortion tensor'' $F_{ij}=\gamma^{1/3} {\cal L}_n \tilde \gamma_{ij}$
(with $\tilde \gamma_{ij}=\gamma^{-1/3} \gamma_{ij}$)\cite{smarryork},
\begin{equation}
D^j D_j \beta^i +\frac{1}{3} D^i D_j \beta^j + R^i_j \beta^j - 2 D_j (\alpha [K^{ij}-\frac{1}{3} K]) = 0 \, ;
\end{equation}
(this result can be also obtained by $D_j (\tilde \gamma^{ij}_{,t}) = 0$).

Recently Brady et. al.\cite{bradyIBBHGAUGE} extended the minimal strain prescription
by minimizing the action with respect to both $\alpha$ and $\beta^i$ obtaining (\ref{ibbhSHIFT}) and
the lapse condition
\begin{equation}
K^{ij} \left ( -2 \alpha K_{ij} + 2 D_i \beta_j \right ) = 0\, .\label{ibbhLAPSE}
\end{equation}
The coupled system (\ref{ibbhSHIFT},\ref{ibbhLAPSE}), is referred to as ``generalized Smarr-York conditions'.
 Recently, Garfinkle et. al. have studied
the question of existence and uniqueness of this system\cite{garfinkleBCT}. The authors
conclude that although there is a potential case for non-uniqueness, this problem
can be avoided by an appropriate choice of slice and boundary conditions.\\

These conditions have the desirable property of reducing the possible distortion
in the spatial coordinates due to the ``evolution'' of the spatial slices\cite{york-sources}.
Additionally, they minimize the rate of change along $(\partial_t)^a$ which is indeed appealing as 
the metric variables should vary slowly in the resulting coordinates. \\

{\bf ``Simplifying'' prescriptions}\\

{\it Coordinate conditions: `Symmetry Seeking Coordinates'} \\
Recently\cite{bradyIBBHGAUGE,garfinkleGAUGE},
prescriptions have been obtained by demanding 
the existence of some `approximate' symmetries. In\cite{garfinkleGAUGE} the authors 
approached the problem by demanding the coordinates be chosen
such that, if the spacetime has an approximate timelike Killing vector, they adapt to the (approximate) symmetry. 
This (pseudo-)symmetry was expressed in terms of a homotetic Killing vector $X^a$, satisfying
\begin{equation}
{\cal L}_X g_{ab} = 2 \sigma g_{ab} \,,
\end{equation} 
(with $\sigma=0$ if $X^a$ is a Killing vector). The homotetic condition
gives rise to evolution equations for $g_{ab}$ which in turn imply equations for $(\gamma_{ij},K_{ij})$;
namely
\begin{eqnarray}
{\cal L}_X \gamma_{ij} = 2 \sigma \gamma_{ij} \, , \label{garfh} \\
{\cal L}_X K_{ij} = \sigma K_{ij} \, . \label{garfk}
\end{eqnarray}
For the coordinate conditions to follow closely the evolutions of the metric variables,
equations (\ref{garfh},\ref{garfk}) are combined with
the evolution equations (\ref{3plus1h},\ref{3plus1k}) to obtain a constrained system (since twelve
equations are obtained but only four variables are to be fixed). There is clearly a vast range 
of possibilities; 
some of the proposed options for the lapse are:
\begin{itemize}
\item{Contraction of (\ref{garfh}) with $K_{ij}$, giving rise to $\alpha=(K^{ij}D_i\beta_j-\sigma K)/(K^{ij}K_{ij})$.
[which will not be useful if $K^{ij}K_{ij}=0$].}
\item{Contraction of (\ref{garfk}) with $\gamma_{ij}$ which results in
$[-D_i D^j \alpha + (R+K^2) \alpha + \beta^i D_i K +\sigma K]=0$.}
\end{itemize}
and for choosing the shift,
\begin{itemize}
\item{Divergence of (\ref{garfh}), resulting in $D^i {\cal L}_X \gamma_{ij}=0$. Which is precisely
the `minimal strain shift' condition.}
\item{Divergence of (\ref{garfk}), which yields $D^i ( {\cal L}_X K_{ij} - \sigma K_{ij}) =0$.}
\end{itemize}
These are elliptic equations and therefore must be supplemented by boundary conditions. Reasonable conditions
for an asymptotically flat spacetime are $\alpha\rightarrow 1$, $\beta^i \rightarrow 0$. Additionally,
inner boundary conditions might be required (for instance in the case of singularity excision). 
These might be specified by Newman or Robin boundary conditions to enforce a $1/r$ behavior\cite{york-sources}. However,
further studies in this direction are needed since, other options might be better suited to `follow'
changes in the dynamics. For instance, in the case of an orbiting system, co-rotating coordinates
should simplify the simulation, and $\beta^i$ at the boundary must be chosen to reflect this fact. (see for
instance\cite{bradyIBBHGAUGE}).\\

Most of the coordinate conditions presented above involve elliptic equations which
might be computationally quite demanding in 3D. In practice, either approximations to these elliptic equations
are used or they are promoted to parabolic equations which are added to the set of evolution equations
under study. \\

{\bf Coordinate Conditions: `Cost-reduced conditions'}\\
Several prescriptions exist that attempt to keep the main properties
of the aforementioned prescriptions while at the same time reducing the 
computational cost of their implementations. Among them,\\

{\it LAPSE}
\begin{itemize}
\item{geodesic slices: Defined by the simple option $\alpha=1$, $\beta^i=0$ (also known
as Gaussian normal coordinates). Although this choice considerably simplifies the equations,
the resulting coordinates tend to converge producing coordinate singularities.}
\item{Harmonic slices: These are defined by $\nabla^a \nabla_a x^b = 0$. This option enlarges
the evolution system with four extra equations and it might lead to coordinate
pathologies\cite{golmCOORDS,hernGAUGE}. However, it has proved quite useful as they help simplify
the evolution equations and been valuable in analytical investigations of the system\cite{york-sources}.
An extension of these conditions, referred to as `generalized harmonic slicing is 
defined by $\nabla^a \nabla_a x^b = F^b$. With $F^b$ a source function chosen to provide more flexibility
and possibly avoid problems encountered with $F^b=0$.}
\item{{\it ``log'' slices:}. This family of slices is introduced by
$d_t \alpha =-f(\alpha) \alpha^2 K$ with $f$ an arbitrary function\cite{arbonaGAUGE}. In particular, for
$f=0,1$ one recovers the geodesical and harmonic slicing conditions respectively.
For the case $f=n/\alpha$ (with $n\in N$), the resulting slicing `mimics' the maximal one
close to large curvature regions (in the sense that the lapse collapses to zero) but in
this case through an evolution equation. } 
\item{{\it ``Evolving'' the elliptic conditions.} In \cite{balakrishnaGAUGE} it is proposed to
promote the elliptic conditions to evolution equations. This idea is basically the way elliptic
equations are solved through an associated parabolic equation. For instance $L(u)=0$ is solved
by considering  instead $\partial_{\lambda}u=\epsilon L(u)$,
with $\lambda$ a relaxation parameter and $\epsilon$ an arbitrary parameter. 
$\lambda$ is chosen to
be the time parameter and the equation for the slice is treated as another evolution equation.
The main disadvantage is that for a stable discretization of the parabolic equation 
a very small timestep might be required (to satisfy the CFL condition\cite{gkobook}) render the implementation
too costly. However, one might choose to relax this equation until some not too severe threshold is satisfied;
the associated cost might be acceptable compared to the one for the elliptic equations as illustrated in\cite{balakrishnaGAUGE}.}
\item{{\it Approximate Maximal Slicings:} The Maximal Slicing equation (\ref{maximalgauge}) for the case
$K=0$ is modified to approximately satisfy this condition, giving rise to a parabolic
equation\cite{shibataGAUGE}.}\\
\item{{\it Slices induced by analytical solutions:} When the system under study is `close' to an
analytically known solution, lapse conditions induced from this solution provide an inexpensive
prescription which can prove useful\cite{usCOLL}.\\}
\end{itemize}

{\it SHIFT}
\begin{itemize}
\item{{\it Pseudo-minimal distortion.} A condition simpler to the minimal distortion is obtained
by replacing the covariant derivative $D_i$ by $\partial_i$. For cases where the spatial
variation of the metric is ``small'', this condition yields a workable approximation
to the minimal distortion shift\cite{nakamuraCOLLNS}. A similar condition is obtained in\cite{shibataGAUGE}
slightly simplified by considering a modification of the action defined in\cite{york-sources}.}
\item{{\it Shift conditions induced by analytical solutions:} Same as slicing condition induced from analytical
solutions.}
\item{{\it Shift conditions tailored for dynamical variable control:} These are conditions derived
by demanding the shift vector be such that some of the dynamical variables are kept constant in time
or driven to a specific value. Having control on the behavior particular variables through 
the evolution can be extremely important. For instance, by demanding that the time derivative of a
particular combination of connection coefficients be `driven' to zero a 
hyperbolic condition is obtained and the overall evolution is notably improved\cite{alcubierreSINGLEBH}.}
\end{itemize}

These coordinate conditions are generic in the sense that they can be applied in any dimension.
For spacetimes with exact symmetries (like spherical symmetry and axisymmetry) further
conditions exist which exploit this property. Particular examples obtained when spherical
coordinates are used are: Polar slices (obtained by enforcing $K=K^r_r$, yielding
a parabolic equation for $\alpha$\cite{bardeen-piran}); Radial or `areal' gauge (so that the area
of surfaces at $r=const$ is exactly $4 \pi r^2$), providing parabolic equations 
for $\beta^i$\cite{bardeen-piran}. \\

\subsubsection{Initial and boundary data\label{initial_data}\\}

{\bf Initial Data}\\
The theory of setting initial data was laid out by Lichnerowicz\cite{lichnerowiczID} and further refined
and expanded by York\cite{yorkID}. (For a recent comprehensive review of the initial data problem and its numerical
implementation refer to\cite{cookREVIEW}). I will here just mention the main aspects of this problem.
The Cauchy initial value problem requires prescribing $\gamma_{ij}$ and $K_{ij}$ on an initial hypersurface. However,
not all these variables are independent. Namely, we know there are {\it four} constraints to be satisfied
and so, only $eight$ out of the twelve in the $\{\gamma,K\}$ pair need be specified. Care must be taken to `single out'
four ``preferred'' variables since under a coordinate transformation the components will mix. This problem
is addressed by the
Lichnerowicz-York's approach which extracts one quantity out of $\gamma_{ij}$ (by 
expressing $\gamma_{ij}=\phi^4 \hat \gamma_{ij}$ in terms of a freely specifiable $\hat \gamma_{ij}$)
and three out of $K_{ij}$ (by expressing the trace-free part of $K_{ij}$ in terms of transverse-traceless
tensor plus a `longitudinal part' which is in turn expressed in terms of a vector $W^{i}$, which
becomes the unknown). An elliptic system of equations for the variables $\{ \phi, W^i \}$ is obtained
that, assuming proper boundary conditions and the freely specifiable data are prescribed, can be solved to
yield consistent initial data to start the evolutionary problem. Of course, the ``free data'' must be given in
such a way to conform to the physical system under study. Spurious radiation on the initial
surface should be minimized and boundary conditions to enforce appropriate 
asymptotic fall-off rates be defined\cite{yorkID}.

Additionally, when dealing with spacetimes containing
singularities, special care must be exercised to handle the singularities. In practice, either the solution is
renormalized, effectively factoring out the divergent part\cite{puncturesID}, or a region containing
the singularity is excised which requires introducing an inner boundary where data must be provided as
well\cite{york-sources,cookID,pedroID_II}. \\

{\bf Inner Boundary Conditions}\\
A particularly delicate issue when dealing with black hole spacetimes is the
presence of singularities. Clearly, a simulation will not be able to handle the
infinities associated with them. In practice, one could use a slicing that
effectively freezes the evolution near  the singularities (like the maximal slicing
condition), but as discussed earlier, the simulation will not proceed for long. Cosmological
censorship\cite{mtw} implies that singularities must be hidden inside the event horizons.
Moreover, the event horizon hides anything inside it; so, in principle, an inner boundary
could be chosen to lie inside the event horizon surrounding the singularity. The presence of
the inner boundary, would prevent the simulations to get `too close' to the singularity
and the simulation should perform well. This idea, originally suggested by Unruh\cite{unruhexcise}
known as {\it singularity excision} is at present the most promising strategy 
to deal with the singularities that might be present in the simulation. There are two basic issues
in implementing this idea: \\
First, since
the concept of event horizon is a global one, it can only
be found {\bf after} the evolution has been carried over. In order to obtain a `local' notion
(ie. on each hypersurface), in practice one looks for trapped surfaces; in particular the outer most
one which is referred to as {\it apparent horizon}. Under certain reasonable conditions, one can
prove that indeed the apparent horizon, if it exists\footnote{Note, 
there is no guarantee that there will be an apparent horizon on any hypersurface, for instance even
Schwarzschild spacetimes admits a, granted odd looking, hypersurface without an apparent horizon\cite{waldNOAPHOR}.
However,
all counter-examples of this type require quite `perverse' looking slices that one can 
`hope' that for reasonable slices one will be found.}, will always lie inside the event horizon\cite{wald}.
Thus, the apparent horizon location is used as a `marker' and the region inside it is excised from
the computational domain, defining an {\it inner boundary} which is either spacelike or null. \\
The second issue, which is a delicate one, has to do with the fact that somehow values at
this boundary must be prescribed. The basic strategy for this is quite simple; since the past 
domain of dependence
at this bounday is `tilted' off this boundary (reflecting the causal structure of the spacetime interior
to the event horizon), one could provide these values using the evolution equations. The
numerical implementation of this strategy, on the other hand, is quite difficult as it must be
capable of dealing with moving boundaries (resulting from singularities moving through the grid);
merging of initially disconnected inner boundaries (like those present in binary black hole spacetimes);
`sudden' appearance of inner boundaries (which would result in collapse situation); etc. An alternative 
way of addressing the assignment of inner boundary values is being developed by Eardley\cite{eardleyIBC,eardleyIBCII}.
This approach explicitly uses the equation determining the apparent horizon (which is assumed define
the inner boundary) supplemented with some geometrically motivated conditions to obtain
a $2D$ elliptic set of equations which can be solved to obtain inner boundary values. (Note: since this approach is
not yet fully developed it has not been attempted so far; but it certainly has appealing properties and should not
be forgotten.)

The numerical implementation of the singularity excision strategy is a delicate issue
and considerable efforts are being
spent in this direction. We will revisit this issue in our discussion of particulars of numerical
implementations (section~\ref{evolcauchy}). \\

{\bf Outer Boundary Conditions}\\
The spacelike slices in `3+1' implementations extend to spacelike infinity $i^o$. 
Assuming, as it is always the case in Numerical Relativity, that the spacetime is globally hyperbolic; data on a given
initial hypersurface completely determines the unique geometry to the future of it. In order
to have a simulation be able to handle these `infinitely large' hypersurfaces, one can in principle, compactify
the spacetime to deal with a `finite domain' and gain access to infinity (where, for instance, the concept of asymptotically
flatness can be used to provide boundary data). 
However, the numerical implementation of this strategy is complicated. Namely, spacetime points
are separated by increasingly larger distances (in particular the boundary point is infinitely
far from the nearest inner neighbor!). As a consequence, there is a clear loss of resolution which considerable
complicates a stability of the scheme. This is a real
problem as ripples in the metric variables ``pile-up" and there can not be enough points to
accurately resolve them. High frequency modes (``noise'') is generated which usually
drives the simulation unstable\footnote{Yet, `noise' that this loss of resolution  creates, could be handled
by carefully filtering them out so as to minimize their influence on the rest of the spacetime.
This approach has been used in\cite{garfinkleCOMPACT} reporting good results for relatively moderate amounts of time.}. 
An approach which has not been pursued yet, is to consider more generic slices, which asymptotatically
become null, that end at future null infinity. In this case,
assuming coordinates have been chosen adapted to the propagation of radiation, the ripples should appear
fairly constant, and the loss of resolution should not be a problem (therefore compactification should be possible).
As future null infinity is approached, terms in the equations tending to $0/0$ will arise,
 which will require special care. Assuming this can be done, 
it would be interesting to see how a `3+1' simulation would proceed when the slices end at ${\cal I}^+$. \\

Because of the potential problems associated with the compactification of spacelike hypersurfaces, 
the most common approach is to {\it ``cut''} the hypersurfaces and bound them with a timelike 
boundary $\Gamma$. Although
 this trivially takes
care of defining a finite  domain for the simulation, it brings about a non trivial one, {\it how to define
appropriate boundary conditions}.
The problem lies in the fact that appropriate boundary conditions are simply not known!. All we know
from analytical studies corresponds to {\it asymptotic fall-off rates} at spacelike or 
null infinity under certain assumptions on the `isolated' source\cite{christodASYMPT,ashtekarASYMPT,gerochASYMPT,friedrichASYMPT}. In
 practice
several strategies are under use:\\

{\it Simplistic approach}\\
The simplest approach is to place the boundaries `as far as possible' and provide data on
$\Gamma$ by simple minded prescriptions
like `freezing' their values; setting them to `educated' guesses on what they should be; etc. Although this approach
provides, at best, approximate values in generic cases, by placing the boundaries far enough from the region one is most
interested in, the error introduced should influence at late times. Hopefully by then, the `interesting' part of the problem
has already happened and one need not worry about the boundaries. This approach clearly is `too dirty' for anyone's taste;
yet, when dealing with simulations that are plagued by instabilities the philosophy has been to try to invest
time improving the treatment of the {\it ``interior''} before the one at the boundary (if, of course
the boundaries are not to blame for the instability, which is a {\bf big if}). Additionally, numerical techniques can
be used to (try to) minimize the reflections; the most commonly used are `filters' 
like {\it sponge filter}\cite{israeliSPONGE,mattMARSA} and {\it blending boundary condition}\cite{gomezBLEND}) which
slightly modify the right hand side of the equations in a `thick' region next to the boundary, where the reflections
are dumped.\\

{\it Radiation Boundary Conditions}\\
A less `crude' approach is to
use the fact that when boundaries are placed in the radiation zone the system must describe
 (neglecting backscattering) purely outgoing waves. This in turn, can be exploited to prescribe approximate boundary
conditions. For instance, imposing Sommerfeld type (outgoing wave) conditions on all variables has been
the preferred choice in most numerical applications, 
(eg. \cite{mattFIRSTCRITICAL,abrahamsBOUND,baumgarteBOUND,nakamuraCOLLNS,shibataAPPHOR}). 
An interesting
option, which has so far not been applied in non-flat spacetimes, is to chose a slicing where the spacelike
surfaces asymptotically approach null ones at the outer boundaries. 
The strategy behind this approach is quite simple, the lapse/shift are chosen in such a way that, asymptotically,
both the hypersurface and lines at constant $x^i$ approach null ones\cite{ethanTHESIS}. The outer 
boundary is effectively ``pushed" further away and the loss
of resolution is not too severe as outgoing fields vary slowly on `close to null' trajectories.
For massless and massive Klein Gordon fields propagating
on a flat background this approach has shown to clearly outperform Sommerfeld type conditions\cite{ethanTHESIS}.
It would be interesting to investigate this strategy in more generic scenarios; with properly chosen coordinate
conditions, this strategy can be really helpful. (Note that providing data on all variables independently is
not consistent as it will be discussed later in this section)\\

{\it Perturbative Boundary Conditions}\\
Boundary conditions have been derived by matching Einstein's equations to a set
of linear equations obtained from linearized perturbations over curved backgrounds\cite{lucianoI,lucianoII}.
This approach neglects the effects of non-linear terms outside the outer
boundary introducing erros which do not decrease with resolution but should become
smaller as the outer boundary is moved futher out. So far, applications of this technique
have been restricted to linear and quasi-linear waves in flat spacetime yielding 
the expected results\cite{lucianoII}. Outgoing waves propagate through the boundaries 
leaving behind a small reflection which can be futher reduced by numerical filtering. \\

{\it Simplistic approach and hyperbolic formulations}\\
The use of strongly/strictly/symmetric hyperbolic formulations
clearly distinguishes the {\it incoming variables} at a given boundary. 
Efforts based on these formulations\cite{spectral2} adopt the standard strategy
of providing `simple minded' or constrained boundary values (see below) but 
in this case {\it only} to the {\it incoming} variables. \\

{\it Constrained Boundaries}\\
There is an important point to be raised here. So far, we have not taken into account that
only {\it two} are the degrees of freedom and imposing boundary conditions to most variables is 
not, in general, consistent.\\
Although in most cases it is difficult to distinguish these two degrees of freedom,
at least we can use that the variables are related by constraints to partially restrict the 
data to provide. For instance if $\Gamma$
is at $x^1=L$, the constraints would be $G_{a1}|_{\Gamma}=0$. Whether these constraints are
satisfied at the boundary by the above prescriptions is not a priori clear. A few studies have been
carried towards specifying boundary conditions satisfying the constraints.

One of them\cite{jeffBOUND} has presented an approach
to incorporate the constraints (induced on a timelike boundary) 
into a 3D ADM evolution code. This work was specifically tailored
for linearized perturbations of flat spacetime and with the shift set to zero; however, this work
evolved the system for about $1000$ crossing times (as opposed to $100$ with Sommerfeld
conditions), showing that
a more consistent approach towards the boundary problem might be quite helpful in a simulation, 
(for a discussion considering a similar approach see \cite{richardITP}).
Another\cite{bardeenPRIVATE}, employs the Hamiltonian and momentum constraint (ie. the constraints
on the spacelike hypersurfaces) to determine boundary values in a 3D code implementing a symmetric hyperbolic
formulation of Einstein equations. Preliminary tests indicate better behaved evolutions are obtained. 
Also, in 1D, constraints have been used to 
provide boundary values and compare with the simple minded approach\cite{spectral2}. For
the case of a Schwarzschild space time, this work illustrates how,
in the tested cases, the prescription of `constrained boundary values' indeed provides stable
implementations while the simplistic approach to freezing incoming field values at the outer boundary fails.

Additional support for the use of constrained boundaries has been presented in the 1D case.
Here a couple of works have chosen boundary conditions defined in a way that
the time derivative of the constraints remain zero (and therefore they are satisfied
throughout the evolution), achieving stable evolution of black hole spacetimes
perturbed with a minimally coupled scalar field without the need of specially designed
gauge conditions\cite{reulaCONS,chicholuisCON}. 

Recently, Stewart presented a systematic study of the well posedness question
of the {\it initial boundary value problem}\cite{stewartIVBP}.
This required analyzing the properties of the evolution system (in this case the symmetric hyperbolic formulation
introduced in\cite{frittellireula}) coupled to the boundary
value specification. He found that well posedness is obtained if these boundary data are specified so that
the constraints are satisfied at the boundary.

Deep insight on the initial boundary value problem both from the mathematical point
of view (ie. well posedness) and its physical interpretation has been presented by
the work by Friedrich and Nagy\cite{friedrichnagy}. Through a careful analysis of the
properties of the system taking into account the presence of a timelike boundary
they conclude that, as expected, only {\it two} variables might be freely specified 
(related to the two polarizations of incoming radiation).
Although the conclusions obtained in this work should be extendible to all formulations (after all it 
is a statement
about the {\it physics} of the problem) the extension is far from straightforward when not dealing
with symmetric hyperbolic systems. Clearly, a more systematic study of the role played by
boundaries in G. R. and their role in numerical implementations is needed. \\

%{\bf A note of caution:} It is worth pointing out here that in a stable implementation
%not providing constrained boundary values, the constraints {\it will be} satisfied (otherwise
%stability can not be achieved). However, the use of constrained valued is likely to aid in 
%the attainment of such an implementation.\\

Another alternative, is to dispense of the outer boundary completely; two options for achieving this are:
Cauchy-characteristic matching\cite{bishopccm1,bishopccm,mattchar,jeffccm,dinvernoEVOL}
or the conformal field equations (see section \ref{form:conformal}).
While the latter implies using a completely different formalism to study the spacetime (and will be presented
in detail in section \ref{form:conformal}), the former 
supplements the `3+1' formulation with a characteristic one (see \ref{form:characteristic}). Basically,
in the region exterior to the boundary to future null infinity, one introduces a foliation along outgoing
characteristics and Einstein equations are written adapted to this foliation.
Since the phase of the `ripples in metric' is nearly constant along  these null surfaces, Penrose's compactification
technique\cite{penrose} is used to deal with a finite computational domain. Just as
several coordinate patches are required to deal with non-trivial topologies, patching together regions
of spacetime treated with different approaches can provide a clean treatment of the problem. \\

The `3+1' approach has been the one receiving the most attention in NR; however, several other
alternatives have been implemented successfully in several systems. These alternatives
are the characteristic
formulation of GR and the conformal Einstein equations.

\subsection{Characteristic Formulation \label{form:characteristic}}

\subsubsection{Formalism:}
The characteristic formulation of G.R. was introduced by Bondi\cite{bondi} and Sachs\cite{sachs}
 in the 60's. The main
strategy of this approach is the use of a foliation by a sequence of (outgoing or incoming) null hypersurfaces
which made it an ideal arena to understand key issues regarding gravitational radiation.
There are several `variants' of this approach yielding slightly different system of equations; however,
they all have in common that only {\it two first order} evolution equations 
and {\it four}  
`hypersurface' equations\footnote{Equations relating quantities only on a given hypersurface.} need be
solved (which are essentially ODE's).
I will here present the one first implemented in 3D\cite{hpgn}, which adopted the Bondi approach to
characteristic GR, but several other efforts have implemented characteristic approaches in $2D$\cite{dinvernoEVOL} or 
$3D$\cite{bartnikALGOR}.

In the Bondi approach a coordinate system adapted to the null foliation is chosen in the
following way: the outgoing (incoming) lightlike hypersurfaces emanating from a
timelike geodesic or worldtube are labeled with
a parameter $u$; each null ray on a specific hypersurface is labeled with $x^A$ $(A=2,3)$ 
and $r$ is introduced as a surface area
coordinate (i.e. surfaces at $r=const$ have area $4
\pi r^2$). In the resulting $x^a=(u,r,x^A)$ coordinates, the metric takes the
Bondi-Sachs form~\cite{bondi,sachs}
\begin{eqnarray}
   ds^2 & = & -\left(e^{2\beta}V/r -r^2h_{AB}U^AU^B\right)du^2
        -2e^{2\beta}dudr \nonumber \\ & & -2r^2 h_{AB}U^Bdudx^A 
         +  r^2h_{AB}dx^Adx^B \, .   \label{eq:bmet}
\end{eqnarray}
Six real field variables appear in this form of the metric\footnote{Note
that the areal $r$ coordinate requirement in turn implies that $\det h_{AB}$ be that of the unit sphere metric;
thus there are only two independent
fields for $h_{AB}$.}: $V$, $\beta$, $U^A$ and $h_{AB}$. They
have a straightforward physical
interpretation: $h_{AB}$ represents the conformal intrinsic geometry of the
surfaces defined by $dr=du=0$ and contains the $2$ degrees of radiative
freedom. The field $\beta$ represents the {\em expansion} of the light rays as they
propagate radially. $V$ is the analog of the Newtonian
potential, and its asymptotic expansion contains the mass aspect of the
system. Note that the coordinate system is tied to null surfaces which can intersect due
to caustics or crossovers. In these cases, the coordinate system becomes
singular! So, it is clear that this approach can not be used for arbitrary
systems. However, as we will discuss in section \ref{causticfix}, one has several options to address
the caustic/crossover problem in a number of cases, thus extending its range of applicability.

The Einstein equations in the vacuum case, $G_{ab}=0=R_{ab}$, decompose into hypersurface
equations, evolution equations and conservation laws.  Bondi designated
as the ``main'' Einstein's equations~\cite{bondi} those which correspond
to the six components of the Ricci tensor, $R_{rr}$, $R_{rA}$ and
$R_{AB}$.

The hypersurface equations, given by $R_{rr}$, $R_{rA}$ and
$h^{AB} R_{AB}$, can be written as
\begin{eqnarray}
\beta_{,r} = \frac{1}{16}rh^{AC}h^{BD}h_{AB,r}h_{CD,r} \: ,
\label{betaori}
\\
(r^4e^{-2\beta}h_{AB}U^B_{,r})_{,r}  =
2r^4  \left(r^{-2}\beta_{,A}\right)_{,r}
-r^2  h^{BC}D_{C}h_{AB,r} \: ,
\label{uori}
\end{eqnarray}
\begin{eqnarray}
2e^{-2\beta}V_{,r} &= {\cal R} - 2 D^{A} D_{A} \beta
-2 D^{A}\beta D_{A}\beta \nonumber \\
 &+ r^{-2} e^{-2\beta} D_{A}(r^4U^A)_{,r}
-\frac{1}{2}r^4e^{-4\beta}h_{AB}U^A_{,r}U^B_{,r}\: ;
\label{vori}
\end{eqnarray}
and the evolution equations, given by $R_{AB}-h_{AB}h^{CD}R_{CD}/2$, are expressed as
\begin{eqnarray}
\fl r {(r {h_{AB}}_{,u})}_{,r} - \frac{1}{2} (r V {h_{AB}}_{,r})_{,r}  &=
 \bigg(2 e^{\beta} D_A D_B e^{\beta} - r^2 h_{AC} D_B U^C_{,r} 
-\frac{r^2}{2} {h_{AB}}_{,r} D_C U^C \nonumber \\
& \mbox{}+ \frac{r^4}{2} e^{-2 \beta} h_{AC} h_{BD} U^C_{,r} U^D_{,r}  - r^2 U^C D_C h_{AB_{,r}} \nonumber \\
& \mbox{}- 2 r h_{AC} D_B U^C + r^2 h_{AC_{,r}} h_{BE} ( D^C U^E - D^E U^C) \bigg) \nonumber \\
&\mbox{} - \frac{1}{2} h_{AB} \bigg( r^2 h^{CD}_{,r} (  {h_{CD}}_{,u} - \frac{V}{2 r} {h_{CD}}_{,r} ) 
+ 2 e^{\beta} D_C D^C e^{\beta}\nonumber \\ 
&\mbox{}-
D_C(r^2 U^C)_{,r} + \frac{1}{2} r^4 e^{-2 \beta} h_{CD} U^C_{,r} U^D_{,r} \bigg)\label{jori} \: ; 
\end{eqnarray}
where $D_A$ is the covariant derivative and ${\cal R}$ the curvature
scalar of the 2-metric $h_{AB}$. There is a natural hierarchy to integrate these equations;
namely, assuming $h_{AB}$ and consistent boundary values are known, the integration
sequence (\ref{betaori})$\rightarrow$(\ref{uori})$\rightarrow$(\ref{vori}), completely determines the metric on a
given hypersurface. Last, equation (\ref{jori}) is integrated to obtain $h_{AB}$ at the next
hypersurface and the process starts again\cite{tamb-win}.

So far, we have accounted for six hypersurface
and evolution equations. Together with the
equations $R^r_a=0$, they form a complete set of components of
the vacuum Einstein's equations. 
Given that the main equations are satisfied, the Bianchi
identities imply they are satisfied on the spacetime provided they hold on a single
spherical cross-section.  By choosing this sphere to be at infinity, Bondi
identified these three equations as conservation conditions for energy
and angular momentum.

\subsubsection{Coordinate conditions.}
It is also possible to obtain geometrical insight into
the fields by analyzing the intrinsic metric of the $r=const$ surfaces,
\begin{equation}
   \gamma_{ij}dx^i dx^j =-e^{2\beta}{V \over r}du^2
        +r^2h_{AB}(dx^A-U^Adu)(dx^B-U^Bdu).
\end{equation}
In analogy to the $3+1$ decomposition of the Cauchy formalism~\cite{mtw}, a
$2+1$ decomposition of the timelike worldtube geometry leads to the
identification of $g_{AB}=r^2h_{AB}$ as the metric of the 2-surfaces
of constant $u$ which foliate the worldtube, $e^{2\beta}V/r$ as the
square of the lapse function and $(-U^A)$ as the shift vector. However,
there is a clear difference. Inspection of the system (\ref{betaori},\ref{uori},\ref{vori})
reveals `hypersurface equations' for the gauge variables; which result from 
the fact that the slices are to be null. As a consequence, the issue of `coordinate freedom' in 
characteristic numerical relativity is not as `open' as in the
Cauchy case, and this freedom is to be fixed at a given timelike or null worldtube.\\

Little has been explored about this choice, most analytical studies have concentrated
on defining the problem at ${\cal I}^{+}$ and integrating the equations radially
inwards. Numerical applications do the opposite, ie. integrate the equation outwards. Additionally,
the remarkable robustness displayed by all characteristic implementations (in the vacuum case) to 
handle superluminal shifts
have not prompted the need to introduce shift choices that would simplify the dynamics. \\

{\it LAPSE}\\
Lapse choices have been induced from analytical solutions\cite{hpgn,bartnikALGOR,dinvernoEVOL,shinkaihayward} or by
matching to a Cauchy evolution\cite{hpgn,dinvernoEVOL}. Additionally, models describing the
geometry of a fissioning white hole have been introduced\cite{fissionwh,husaTORUS} in which
the parametrization of the null generators can be used to induce lapse conditions for a double
null evolution\cite{jeffPROGRESS}.\\

{\it SHIFT}\\
Although vacuum codes routinely handle superluminal shifts without problems, simulations of 
non-vacuum systems\cite{nigelmatter,philipMATTER}
might benefit from a convenient choice. For instance, when modeling a `star' orbiting
around a black hole, a shift can be used so that the angular coordinates rotate around the inner boundary
``following'' the orbiting star which, in the resulting coordinates, will
remain (approximately) fixed\cite{nigelNEW}.

\subsubsection{Initial and Boundary data \\}

{\bf Initial data}\\
A distinctive feature of the initial data problem in the characteristic formulation is that
data on a given initial hypersurface are generally not enough to determine the solution (not even locally). 
This is due to the fact that the domain of dependence of a single {\it nonsingular} null hypersurface
is empty!. In order to obtain a well defined problem the null hypersurface must either be completed
to a caustic-crossover region or an additional boundary must be introduced (which defines an
$S^2$ cross section at the intersection). In present numerical
applications the latter option is pursued where the boundary is either null or timelike.
Assuming the constraints are satisfied in this inner boundary at the $S^2$ intersection, one can freely
chose $h_{AB}$ on a given surface (albeit subject to a regularity conditions at the intersection), integration of
the ordinary differential hypersurface equations yields a perfectly valid initial data {\it without}
having to solve an elliptic problem. The non-elliptic character of these equations is a consequence
on their application on a null surface, rather than spacelike. For the case the boundary is null, the system
is well posed\cite{rendalCHARACT}; for the timelike case, only existence and
uniqueness has been proven\cite{zumhagen,fritlehn}.

Although there is no difficulty in obtaining `valid' initial data, the important issue is to have this
data be `physically relevant'. Cauchy formulations can reach to Post Newtonian approximations for guidance
in the search for physically relevant data, in the characteristic case, an approximation approach based on
a family of null cones with the speed of light being a varying parameter\cite{jeffnewtonian} has been
introduced to make contact with Newtonian theory. This approach guarantees that for weakly radiating
systems the obtained waveforms are, to first order approximation, given by the quadrupole formula.\\

{\bf Boundary data: Inner boundary}\\
In implementations, when the inner boundary is timelike, the data have been defined by either known
analytical solutions\cite{eth,cce,shinkaihayward} or through matching  to a 3+1 evolution
being carried out in the interior (we will discuss more on matching 
in section~\ref{causticfix})\cite{matching,dinvernoCCM}. These 
options guarantee
the extra four equations ($R^r_a=0$) are satisfied at the boundary. 
In the case the inner boundary is null, since whichever data have been given on the initial null hypersurface
can not interact with the boundary, these can be easily specified. In particular applications,
 the inner boundary
has been chosen to coincide with the past null horizon of a Schwarzschild spacetime\cite{hpgn,mattchar,bartnikALGOR};
an incoming null surface (outside the event horizon) of a Kerr spacetime\cite{shinkaihayward}
or in a double null problem where the inner boundary corresponds to a fissioning white hole (this case will be later
discussed in more detail in section~\ref{causticfix}).\\

{\bf Boundary data: Outer boundary}\\
Another property that makes this formulation appealing is that the outer boundary is
${\cal I}^+$, the hypersurfaces define
cuts at ${\cal I}^+$ which is a flat $S^2 \times R$ null manifold, defined by the end points
of outgoing null curves. No boundary condition is needed as the evolution
proceeds along ${\cal I}^+$ at this boundary.
Since gravitational waves have constant phase on null hypersurfaces
the compactified spacetime can be safely implemented numerically without the risk of the loss of
resolution affecting the evolutions. Additionally, having access to future null infinity brings about
extra benefits, like the possibility of rigorously obtaining the gravitational radiation, mass and
angular momentum\cite{bondi,sachs,tamb-win,geroch-winicour}; also, when studying asymptotically 
flat spacetimes, the metric variables
have a well known asymptotic dependence which has been exploited to aid the numerical
implementations\cite{hpgn,dinvernoEVOL,bartnikALGOR}. \\

\subsection{Conformal Einstein Equations \label{form:conformal}} 

\subsubsection{Formalism:}
A further approach used in numerical relativity is known as the `conformal Einstein equations approach' and was
introduced by Friedrich in the early 80's\cite{friedrichORIG}. The main peculiarity of this approach is that instead of 
solving for the spacetime $(M, g_{ab})$, it first obtains the description of a larger one $(\tilde M, \tilde g_{ab})$.
As a result, one can  foliate the spacetime  $\tilde M$ with a sequence of spacelike; null or more generic 
hypersurfaces. Although the latter option has not been pursued to date, the former approach has been adopted
in all efforts. Naturally, this approach is also of Cauchy type
but I have chosen to present it separately as it has a few notable differences with those from section~\ref{form:cauchy}.
The larger spacetime is determined by the conformal Einstein equations which can be
expressed as
\begin{eqnarray}
\label{Riccigleichung}
\fl \tilde \nabla_a {\tilde  R}_{bc} - \tilde \nabla_b {\tilde R}_{ac}
+ \frac{1}{12} \left( (\tilde \nabla_a \tilde R) \, \tilde g_{bc} - (\tilde \nabla_b \tilde R) \, \tilde g_{ac} \right)
+ 2 \, (\tilde \nabla_d \Omega) \, d_{abc}{}^d & = 0, \; \; \\
\label{Weylgleichung}
\tilde \nabla_d d_{abc}{}^d & = 0, \; \;\\
\label{OmGl}
\fl\tilde \nabla_a \tilde \nabla_b \Omega_a + \frac{1}{2} \, {\tilde  R}_{ab} \, \Omega 
- \frac{1}{4} \tilde \nabla^a \tilde \nabla_a\Omega \, \tilde g_{ab} & = 0, \; \;\\
\label{omWllngl}
\fl\frac{1}{4} \tilde \nabla_a \left(\tilde \nabla^b \tilde \nabla_b \Omega\right) + 
\frac{1}{2} \, {\tilde  R}_{ab} \, \tilde \nabla^b \Omega 
+ \frac{1}{24} \, \Omega \, \tilde \nabla_a \tilde R + \frac{1}{12} \, \tilde \nabla_a
\Omega \, R 
& = 0,\; \; \\
\label{irrRiemann}
\fl  \Omega d_{abc}{}^d
  + ( \tilde g_{c[a} {\tilde  R}_{b]} {}^d
  - \tilde g^d{}_{[a} {\tilde  R}_{b]c} ) 
  + ( \tilde g_{c[a} \tilde g_b]{}^d ) \frac{\tilde R}{6} - \tilde R_{abc}{}^d & =
  0,\; \; \\
%
%
%\noalign{and}
\fl\label{Rtrace}
  \Omega^2 \tilde R + 6 \, \Omega \, \tilde \nabla^a \tilde \nabla_a \Omega -
    12 \, (\tilde \nabla^a \Omega) \, (\tilde \nabla_a \Omega) & = 0. \; \; 
\end{eqnarray}
A solution of this system provides the metric $\tilde g_{ab}$ (defining a unique covariant
derivative $\tilde \nabla_a$),
the traceless part of the Ricci tensor ${\tilde R}_{ab}$, the Weyl tensor
(of $\tilde  g_{ab}$) $\Omega d_{abc}{}^d$ and $\tilde R$ (the Ricci scalar).
The physical spacetime $M (\subset \tilde M)$ is defined by $M:=\{p \in \tilde M | \Omega > 0\}$
($\Omega=0$ represents the boundary of $M$).
The metric $g_{ab}:=\Omega^{-2} \tilde g_{ab}$ is a solution of Einstein equations
on $M$. It is worth pointing out that the (degenerate) physical metric at $\Omega=0$ is also obtained,
thus, one straightforwardly gains access to future (or past) null infinity and
quantities like gravitational radiation and tidal forces at infinity are obtained
by straightforward algebraic evaluations.
Although this system seems more complex, it is also amenable to a sort of $3+1$ decomposition\cite{friedrich96}
in much the same vein as that presented in section~\ref{form:cauchy}. 
$\tilde M$ is sliced with a parametrized (with parameter $t$)
sequence of spacelike hypersurfaces $\Sigma_t$. The unit normal to $\Sigma_t$ given by $n^a$, allows for adopting
the intrinsic and extrinsic curvatures of $\Sigma_t$, denoted by $h_{ab}$ and $K_{ab}$ as main variables. 
Additional variables are
introduced to reexpress the system
in first order form and obtain a symmetric hyperbolic system of equations for the variables
$(h_{ab},K_{ab},\gamma^{a}_{bc},\Omega,\Omega_0,\Omega_a,\omega,E_{ab},B_{ab},R_a^*,R_{ab}^*)$. Here
$\gamma^{a}_{bc}$  is the 3-connection of $h_{ab}$; $E_{ab}$ and $B_{ab}$ are the electric
and magnetic parts of $d_{abcd}$; $\Omega_o=n^a \tilde \nabla_a \Omega$; $\Omega_a=h_a^b \tilde \nabla_b \Omega$
and $R_{ab}^*$, $R_a^*$ are particular projections of $\tilde R_{ab}$.
Clearly, the system contains many more variables than the traditional ADM approach. However,
its is important to point out that: (i) some of the variables are directly related to the gravitational
radiation (and there is no extra work to obtain it from the evolved data) and (ii) the system is well posed,
and the number of variables is certainly comparable to (most) well posed formulations obtained in the traditional
`3+1' approach.
\\
Aside from the `standard gauge freedom' described by the lapse and shift vector,
there is a further one in any conformal approach. Note that the conformal and the physical metric 
are related by a rescaling which is essentially arbitrary, as two solutions $(\tilde M,\tilde g_{ab},\Omega)$ and
$(\tilde M,\bar{g}_{ab},\bar{\Omega})$ with $(\bar{g}_{ab},\bar{\Omega}) =
(\theta^2 \tilde g_{ab},\theta \Omega)$ and a positive function $\theta$
describe the same physical spacetime. Under the rescaling $\theta$, the Ricci scalar $R$ changes. Specifying either
$\Omega$ or $R$ fixes this freedom.

I have presented the conformal equations in the 3-tensor formalism simply because 
it is the one that yielded a $3D$ implementation and its `closeness' with
the `3+1' presentation of section~\ref{form:cauchy}. However, the equations have also been 
presented in the spinorial~\cite{frauendienerEVOL} 
or frame formalisms~\cite{friedrichORIG,peterFIRST}. 

\subsubsection{Coordinate conditions.}
Choosing gauge conditions for the conformal equations is a similar problem to the `3+1' approach. Care in this
case must be taken so that the foliation crosses ${\cal I}^+$ and not ${\cal I}^-$ as one tries to avoid 
going trough $i_o$ (among other reasons so that boundary conditions on the unphysical spacetime will not
propagate into the physical one).\\

{\it LAPSE}\\
The options for the lapse used so far have been obtained from analytical expressions;
derived from harmonic conditions\cite{frauendienerEVOL}
or from the condition $\alpha = e^{s} \sqrt{det(h_{ab})}$ (with $s$ real)\cite{peterLAST}. 
At first sight, this last 
condition appears awkward as it would suggest that evolution is `accelerated' when $det(h_{ab})$ becomes large.
In simulations of Schwarzschild spacetimes\cite{peterITP}, this has not represented major
difficulties since the initial slice is chosen to be `far' from the singularity. 
Nevertheless, as more generic
initial data is considered, the need for alternatives for the lapse would likely be greater. \\

{\it SHIFT}\\
A particularly interesting choice for the shift, is one which keeps the location of ${\cal I}^+$ at a constant grid
location\cite{frauendienerEVOL}. This addresses a common criticism to this formulation where future null infinity
can move inwards in the
grid and therefore, computational resources are wasted more and more since the unphysical space becomes
larger (with respect to the grid). This choice introduced by Frauendiner has been successfully implemented
in $2D$ to study vacuum spacetimes with toroidal null infinities and read-off the gravitational radiation
at ${\cal I}^+$\cite{frauendienerEVOL}.\\

As mentioned previously,
numerical implementations have also been presented in the {\it frame} 
formalism~\cite{friedrichORIG,peterFIRST},
which can be more `flexible' with respect to gauge choices. \\

\subsubsection{Initial and Boundary Data\\}
{\bf Initial Data}\\
The literature on choosing initial data is not as extensive as in the traditional `3+1' approach
as the numerical implementation of the conformal approach is considerable `more recent' in time.
 However, the picture does resemble the `3+1'
approach as constraint equations limit the possible configurations of the initial data
$(h_{ab},K_{ab},\gamma^{a}_{bc},\Omega,\Omega_0,\Omega_a,\omega,E_{ab},B_{ab},R_a^*,R_{ab}^*)$.
As proven in\cite{initdatafriedrich,initdatafriedrichII}, only a subset of data need be solved,
namely by solving an elliptic system for $(h_{ab}, K_{ab},\Omega, \Omega_0)$ simple contractions 
on the remaining constraints yield the complete set of variables. Hence, the initial data problem,
at least from the elliptic system to be solved, is by no means more complicated than the one in
the traditional system. In fact, it would be reasonable to assume that much of the numerical expertise gained to solve
the traditional system  should be `transferable' to the conformal approach. \\

{\bf Inner Boundary data}\\
Just as in the previous formulations, if the hypersurfaces contain singularities one can use singularity
excision techniques to excise the singularities from the computational domain as was done for
the $1D$ scalar field collapse presented in~\cite{peterFIRST}. Another option which in fact has
been the preferred one in the 3D simulations of Schwarzschild spacetime\cite{peterEVOL} has been to 
use slices that do not contain
the singularity; namely the foliation was chosen so that the slices cross 
both ${\cal I}^+$s of the Kruskal extension\cite{wald} of the Schwarzschild
spacetime without ``hitting'' the singularities. Clearly, this approach
 is sound and could be also used in the `3+1' approach (assuming variables can be properly renormalized
at $i^o$ or ${\cal I}^+$, the latter case being more or less straightforward in the conformal approach); however,
 the simulation is making roughly twice the work
(there is no need to evolve sector IV in the notation of\cite{wald}). Moreover, it is not clear whether
a spacetime with two black holes would  be amenable to such strategy since the gauge conditions will have to be carefully
tuned so that the slices avoid both singularities. 

Singularity excision would seem to be better adapted to handle more generic situations. Incorporating
this technique to the conformal approach should be expedited by the expertise (being) 
gained in this area in the `3+1' approach. \\

{\bf Outer Boundary data}\\
The spacetime under study in this case is larger than the physical spacetime. As a consequence,
the outer boundary lies beyond future null infinity. At first sight
it would appear awkward to set up conditions at this boundary since it is not known
what boundary conditions are to be specified there. However, this need not be a problem, since ${\cal I}^+$ is
an incoming null surface and the space beyond ${\cal I}^+$ is causally disconnected from the
physical spacetime. Thus, this formulation manages to get rid of the boundary problem by `hiding' 
the boundary from the region of interest.  There is a price to pay for this feature, namely that 
the implementation spends time evolving points that are of no interest and there is therefore extra
computational overhead. In principle, this can be minimized by adopting an appropriate shift
conditions\cite{frauendienerEVOL} that keeps the location of ${\cal I}^+$ at a constant coordinate value. \\

\section{Some ado about numerics}\label{muchado}
Now, suppose one has (i) decided for a given system of equations for a set of variables;
(ii) adopted suitable coordinates and/or coordinates conditions  and (iii) defined the equations
which determine the initial and boundary data and feels ``ready'' to implement (i-iii) numerically.
The question to ask is: {\it How does one proceed to obtain such implementation?}

First, a ``finite"  representation of the (continuous) (n-1)-dimensional hypersurfaces
is obtained by defining a (not necessarily uniform) grid or lattice whose vertices
can be labeled by a discrete set of points $(x^1_{i_1}...x^{n-1}_{i_{n-1}})$ (with $i_j=1..N_j$). Then, 
a finite representation
for the field variables is obtained by either (I) representing the variable by its value 
at points in the grid  $\Psi^n_{i_1..i_{n-1}}\equiv\Psi(t^n,x^1_{i_1}...x^{n-1}_{i_{n-1}})$ 
or (II) expanding the variable on a finite set of trial 
functions; ie $\Psi(t^n,x^1...x^{n-1})=\Sigma_l^N C_l^n \phi_l(x^1...x^{n-1})$. The finite representation 
is then given by values of the variables 
themselves, $\{\Psi^n_{i_1..i_{n-1}}\}$ (case I) or the
coefficients $\{C_l^n\}$ (case II).

These two different strategies yield, as expected, two very different approaches. {\it Finite difference
approximations} belong to case (I), while {\it Finite difference
elements}; {\it spectral methods}; {\it multiquadrics}, among others, belong to case (II).
Irrespective of the method used, the `end' result is an algebraic problem, which, in the limit
of infinite resolution (ie. grid points spacing$\rightarrow 0$, for case (I) or
$N\rightarrow\infty$ for case (II) ) the algebraic system should
 reduce to the original PDE system\footnote[1]{
This is known as a {\it consistency} requirement; although I would prefer the term {\it absolute}
condition, since otherwise one is not studying the system of interest!}.

\subsection{FDA: A couple of useful points}
Finite difference approximations (FDA) are widely used 
in computational physics and are so far the most popular choice in numerical 
relativity. The details of this technique can be found in most numerical
analysis books (for instance~\cite{recipes,thomas,gkobook}); I will here comment
on two important points which are not often discussed.

A finite difference approximation (FDA) entails replacing all derivatives operators
by discretized counterparts. These discrete operators approximate the derivative
of functions using the grid values $\{\Psi^n_{i_1..i_{n-1}}\}$ and can be obtained
through formal Taylor expansions. There are an infinite number of combinations
that a priori can be used to approximate the original system. Unfortunately,
the majority of these combinations result in unstable implementations. This is often reflected
in the high frequency components of the solution growing without bounds. In practice,
stable implementations often  ``control'' this potential problem by dissipating the high frequency modes.
In nonlinear systems, this proves to be very important since, even when the initial data do not
contain high frequency modes, these will likely be generated by the low frequency ones.

Extensive analysis of dissipative schemes to obtain stable discretization of wave equations
was performed by Kreiss and Oliger\cite{kreissoliger}. They showed how the addition of 
dissipation could become crucial when treating nonlinear systems. The value of such techniques
have been validated over the past fifty years since they were first proposed by
Von Neuman and Richtmeyer\cite{neumanDISSIP} to solve the classical Euler equations. In numerical relativity
their use can be traced back to Wilson's implementation of the relativistic hydrodynamic equations\cite{wilsonII}.
In more recent times, dissipation techniques have been shown to be of 
great help in achieving stable discretizations in computational relativity, 
for instance in\cite{mattFIRSTCRITICAL,luisJCP,teukolskyNSCOLLISION}.

Additionally, the use of dissipation can play a crucial role for achieving stable discretizations
for initial boundary value problems. This is highlighted in the work by Oliger\cite{oligerBOUNDARY}
who considers the equation 
\begin{equation}
F_{,t}=a F_{,x} + b(x,t) \, \label{1dwave}
\end{equation}
 in the domain $[L_1,\infty)$ where inner boundary conditions at $L_1$ are expressed as
\begin{equation}
F^{n+1}_{ib} = \Sigma_{k=0}^m A_k F^n_{ib+k} + g^n_{k} \label{approxbounda} \, ;
\end{equation}
with $m$ indicating the number of points to the right of $x_{ib}=L_1$ involved in the scheme.
For instance, a particular case of eq. (\ref{approxbounda}) would be
\begin{equation}
F^{n+1}_{ib} = F^n_{ib} + \frac{\Delta t}{\Delta x} ( F^n_{ib+1}-F^n_{ib} ) \, .
\end{equation}
Oliger proved the following theorem\cite{oligerBOUNDARY}: {\it If the approximations for the
initial value problem and for the approximation at the boundary (\ref{approxbounda}) are {\bf stable} 
and, further, (\ref{approxbounda}) is {\bf dissipative} then, the implementation of the initial boundary value problem
is stable.} \\

This result shows the following: (I) Stability of the initial boundary value problem
can be assessed by providing boundary conditions written in PDE-like form.
(II) The stability and dissipative properties of this equation can be
readily obtained which coupled to the stability of the initial value
problem provide a stable implementation.

Naturally, it would be desirable to have similar results tailored to the more
complicated systems considered in Numerical Relativity. I doubt this will be achieved 
since the non-linearities and coupling of Einstein equations make a similar
analysis quite difficult. Nevertheless, as we will
see later in section \ref{evolcauchy}, the equations are customarily recast in a form somewhat closely related to
eqn. (\ref{1dwave}) and it is important to keep this theorem in mind. The use of dissipative
inner boundary conditions has not yet been generally pursued; however, I am aware of
a few systematic efforts in this direction reporting considerable
improvements\cite{golmEXCISION,mattGRAXI,lehnerDISS}. The advantages gained from the use of
dissipation both in the absence and presence of boundaries indicates that implementations
can benefit considerably from its use.\\

As mentioned, FDA have been the preferred choice in Numerical Relativity,
their ease of use; transparent interpretation of its strategy and power certainly make
them very attractive. This is illustrated by their use in all areas of numerical relativity; ie.
initial data problem, evolution and ``physics'' extraction.
There are a few criticisms which have lead people into
other choices,
\begin{itemize}
\item{Appropriateness of its use on arbitrary variables: Basically, when using Taylor expansions up to
order {\it n}, one {\it exactly} accommodates for polynomials up to the n-th order. However,
this might not be the ideal `basis' to express certain functions at particular places. For instance $1/x$
near $x=0$ is not conveniently represented by polynomials (of positive integer). 
A solution to this problem is to reexpress variables so that they are better represented
by polynomials; thus if a given function $F$ is expected to behave like $1/x$,
reexpressing the equations in terms of a variable $\tilde F = x F$ improves the
obtained results. This technique has been used in a limited number of
cases\cite{puncturesID,cce,garfinkleCOMPACT,bartnikALGOR} yielding excellent results. However, this approach
requires some `prior' knowledge of the fields dependence.}
\item{Awkward use at non-regular boundaries: As discussed, the variables are represented
by their values at grid points; when dealing with irregular boundaries where
values and derivatives might be required, an often complicated set of interpolations must be carried out. This
introduces high frequency modes which brings about all sort of nightmares. Dissipation
of these modes could take care of this problem but requires carefully designed algorithms. 
As the grid is refined, this problem might become less severe. Refining a grid (ie. adding more points to it)
increases the computational cost considerably; however, the use of {\it adaptive mesh refinement}
can help to alleviate this problem by refining the grid locally only where needed (more on this
technique in section~\ref{helpingnumerics}).}
\end{itemize}

Certainly, these criticisms can be addressed but, undeniably, some 
situations might be better handled by other methods. For instance, expansion
in terms of spherical harmonics of a regular enough variable, say the electromagnetic
potential of localized distribution of charges, might yield an accurate
and inexpensive representation nicely adapted to a particular problem. In cases like this, 
the use of appropriately chosen basis functions are of great help. There are several approaches
based on this idea being used in Numerical Relativity\cite{lagunaMETHODS} and I next
briefly review some of them. \\

\subsection{Beyond FDA}

\subsubsection{Finite Elements.}
The use of Finite Elements (FE) in Numerical Relativity has so far been restricted to 
the solution of the initial value problem of the `3+1'.
Here, the flexibility of this approximation to conform to non-regular boundaries is
a valuable asset. Namely, the `discretized' version
of the hypersurface constitutes a `mesh' of, usually, triangles which
are not required to be regular. As a consequence, hypersurfaces with `holes' are conveniently
covered (which is often more difficult with FDA). Additionally, if steep gradients are expected, 
smaller sized triangles can be used to 
accurately represent them. These particularly nice features come at a price, as the solution
is obtained through a global minimization of the `residual'.  Roughly, the solution $S$  of
the equation $L(S)=0$ is approximated by 
\begin{equation}
\hat S(\vec x) = \Sigma_l^N a_i \phi_s(\vec x) \, ;
\end{equation}
where $a_i$ are unknown coefficients and $\phi_i(\vec x)$ known basis functions (which
are continuously differentiable and integrable functions).
The numerical implementation will not, in general, exactly satisfy the original equation
but $\hat L (\hat S) = \hat R$ (with $\hat L$ the discretized version
of $L$ and $\hat R$ the residual). By minimizing $\hat R$ on the {\it whole} computational domain
an algebraic system for $\{a_i\}$ is obtained. This method has a `global' flavor nicely suited
to the treatment of elliptic equations. Its flexibility to treat irregular boundaries
has been implemented to solve the initial value problem of Einstein equations in\cite{arnold}
where multigrid techniques have been used to diminish in part the high computational cost. \\

\subsubsection{Spectral Methods.} 
Other interesting options  are
the spectral and pseudo-spectral methods\cite{canuto,boyd}. Not only have they been used for
the initial data problem\cite{spectral0,peterID,frauendienerID,gourgollom,bonazzola2BHS}, but
 are being employed for the actual evolution
part\cite{spectral1,bartnikALGOR,generalEC}. These methods have the capability of addressing
non-trivial boundaries without the overhead required for a minimization procedure
(although the goal is to minimize the residual error, as in Finite Elements methods, this is done only at
particular  {\it collocation} points conveniently distributed on the computational domain). 
In this method, the solution is expanded in terms of a set of basis functions
(usually trigonometric functions or Chebyshev polynomials). In spectral methods,
the PDE system is Fourier transformed 
to obtain a simpler one in the frequency space whose solution
is then transformed back to produce that of the original system. Depending on the type of
PDE under study, this transformation might not yield a simpler system in the frequency space.
For these cases, Pseudo-spectral methods were introduced. Loosely speaking, in these methods
only part of the system is treated in the frequency space while the other is solved in the
coordinate domain (for instance time derivatives are done in the regular space while spatial ones
in the frequency space).
The Fourier transformation is in practice carried out in an efficient way through the use of
fast-Fourier transformations. For problems {\it with smooth solutions} these methods
converge {\it exponentially} as the number of basis functions is increased. This improved
convergence rate comes at a higher computational cost, which is nevertheless justified.
Two problems are often cited as the main ones. First, in evolutionary problems, the CFL condition\cite{kreissBOOK}
(which requires the numerical domain of dependence to contain the analytical one) scales as
$N^{-2}$ (while in general FDA scales as $N^{-1}$) which can render the application too costly (note however
that for smooth functions small values of $N$ are usually enough).
The second problem relates to the way the collocation points are chosen which requires the computational
domain be sufficiently simple. This is a problem when dealing with a spacetime containing
irregular boundaries like those containing more than one black hole. It has been suggested that the use of 
several overlapping regions (known as domain decomposition) can overcome
this problem\cite{peterID,spectral1}; and the solutions on each patch would serve as boundary conditions
for the other patches. The scheme would involve an iterative procedure which would, hopefully, converge.
This suggestion is justified by the fact that this strategy indeed works for the Laplace
equation\cite{quarteronivalli,gourgollom}. Considerable progress has been obtained with Einstein equations
and the obtained results are so far very good\cite{generalEC,bonazzola2BHS}.\\

\subsubsection{Latticed based approaches: Regge Calculus and Smooth Lattice method}
In the 60's Regge introduced a way of approaching General Relativity which by its
discrete nature appeared tailored for Numerical Relativity. Rooted in the ADM formalism,
it replaces the dynamical field variables by finite distances by the following approach.
A lattice is introduced and the main variables correspond to the length
of (short) geodesic segments defining the legs of the lattice. A related approach
implements the ADM equations directly on the lattice introducing a series of local Riemann
coordinates\cite{reggeBREWIN}. To date the application of lattice based approaches have
been rather limited; mainly to model the Kasner $T^3$ cosmology
and Schwarzschild spacetime\cite{gentlemiller2,reggeBREWIN}.
A project to investigate this approach in
more general scenarios is under way\cite{gentlemiller}. Initial data corresponding to 
gravitational waves on Minkowski and Shwarszchild backgrounds and
head-on binary black holes (Misner data) have been obtained tailored for a code 
implementing Regge calculus. We should soon
hear reports on the feasibility of this approach to study generic settings.\\

\subsection{Simulation costs and how to improve the picture: AMR, Multigrid,
Parallelism.\label{helpingnumerics}}
Let us estimate the computational cost $CC$ to carry a $3D$ simulation, say for instance
we want to model a black hole system. To fix
ideas let's assume we will employ FDA and the ADM formulation on a uniform grid
with $N_p$ grid points in each direction.
The number of operations needed to `advance' a single time step will be
given by $N_p^{3} \times$~number of floating point operations
per point ($CC_1$). A back of the envelope estimate for the operations is:
Number of variables $\times$ Number of operations per variable $\times$ Number of
`updates' per timestep (eg., if we are using predictor-corrector types of algorithms,
this last item would at least be $2-3$).\\

NU $\equiv$ Number of `updates': $3$.\\

NO $\equiv$ Number of operations: the Ricci tensor appears in the rhs of the equation 
and its evaluation requires $\approx 2000$ floating point operations.\\

NV $\equiv$ Number of variables: $12$ (from $\{g,K\}$) $+$ $4$ (lapse and shift) $+$ $1$ (marking
variable to keep track of where the holes are at each step)). (these have to be multiplied
by $2$ to keep the `old' and `new' values). Hence we have on the order of $\approx 30$ variables.
Thus
\begin{equation}
CC_1 \approx  2 \, 10^5 \frac{NO}{2000} \times \frac{NV}{30} \times \frac{NU}{3} \, .
\end{equation}
Now, suppose the typical size of the source we wish to include in our simulation is $M$.
We must be capable of placing the outer boundary in the wave zone, which would
require our computational domain be at least $[-20M,20M]$. The resolution to (barely) resolve
the system will be $\Delta x = M/4$. Hence $N_p \approx 160$.
In order
to resolve the first quasinormal modes of the produced radiation, we would like the total
simulation length be $\ge 100M$. Since, stability requirements would imply (assuming a
fully explicit FDA approximation) $\Delta t \approx \Delta x/4$ the total number of
timesteps required is at least $N_T=10 N_p$. Therefore the total computational cost would
be $CC = 10 CC_1 N_p^4$.
\begin{equation}
CC \approx 2 \, 10^{14} \frac{NO}{2000} \times \frac{NV}{30} \times \frac{NU}{3} \times
\left( \frac{N_p}{160} \right)^4  
\end{equation}
The most powerful publicly available chips nowadays have peak performance of $10^{9}$ 
floating point operations per second (1 Gflop/sec). Therefore, our full 3D simulation would take
$10^5 secs. \approx 30 hrs$. Memory wise, every real number is at least represented by $R_p=8$ bytes. 
Since
one usually introduces temporary variables to aid in the calculation, in practice,
the total number of variables from the previous estimate at least doubles, so, the 
memory requirements ($MR$) would be
\begin{equation}
MR \approx 2 \, 10^9 \,  \left( \frac{N_p}{160} \right)^4 \times \frac{NV}{30} \times 
\frac{R_p}{8}
\end{equation}

These numbers are not too bad, but are to be considered as an ``idealized lower bound'' since
we have considered the minimal required configuration in the vacuum case (for spacetimes containing
fluids, $\Delta x$ is usually required to be much smaller or the dynamics of the fluid will not be
represented accurately). Additionally, many simulations will be needed for a reliable configuration
space survey and the total computational time invested will increase considerably.
Moreover, if we wanted to perform the same simulations with a better resolution, things rapidly
increase. For instance, improving our resolution by a factor of $2$ would increase
$CC$ by a factor of $16$ and $MR$ by 8 (ie. now we would have to wait 20 days
for the results and need 8 times more memory). The computational cost of 
symmetric hyperbolic formulations would be of about the same order (more expensive though)
but the one for the characteristic
simulation much less (in this case, arrays need only 2D storage and the right hand side
require about $\approx 200$ evaluations).

As we have seen, the computational cost to go beyond the `bare necessities' of
a simulation in 3D rapidly increases. However, there are computational techniques 
that allow finer resolved simulations be achieved without paying such a high price.
I will next mention a few of those.\\

{\bf Adaptive Mesh Refinement}\\
When modeling systems, like gravitational collapse, black hole/neutron star spacetimes,
singularity structure, etc.; the
strength and variability of the field variables are expected to be significant only 
at a `small' region. Achieving an
accurate model capable of capturing the essential features of the dynamics might
require keeping the local truncation error below some threshold. In practical terms,
this often requires much more information from the variables in these `small' regions.
Clearly, one can adjust the overall resolution by satisfying the strongest requirement and therefore
 enough information will be available for
all regions. This straightforward approach is evidently sound; however, 
it might entail wasting computational resources in regions where not much is `going on'.
A more desirable strategy would be to choose a non-uniform grid or definition of collocation
points adapted to those regions that need to be resolved better. Here we again face the problem
that in general we might not know this a priori! One could, in principle, proceed with coarse grid
first, and from the obtained solution deduce properties that a subsequent finer one 
should have. This strategy has as weakness that the `coarse' solution might be too crude to
produce a good enough solution from which to infer how to proceed. If this is indeed the case,
one could discard the ``coarse" simulation and start all over with a finer one. \\

A more direct approach, and one that in principle should work directly (ie. without trial and error)
is to `adaptively' increase or decrease the information needed locally by monitoring the
solution `on the fly'.
In computational relativity,  this approach has so far only been used in simulations using FDA and is known as {\it adaptive
mesh refinement}. This method adds more points to the grid according to some user-defined
threshold on the local truncation error. The use of adaptive mesh refinement in 3D numerical
relativity is making its first steps\cite{brugmanAMR,seidelAMR,dienerAMR,centrellaAMR}, but its benefits
have been dramatically confirmed
by the investigations of Choptuik in 1D\cite{mattFIRSTCRITICAL}. Choptuik employed
a technique introduced by Berger and Oliger\cite{bergeroliger}
to write a fully adaptive code to solve the Einstein-Klein-Gordon system in spherical
symmetry. This allowed him achieving very high accuracy with relatively low computational
cost, and more importantly, to discover critical phenomena in G.R.
Today, computational speed and memory resources are readily available for very fine 1D 
simulations without the use of AMR. However, in 3D where
one barely has enough resources to achieve crude simulations, the use of AMR would open
the door to better resolved simulations, and perhaps, many of the nightmares faced by numerical
relativity in 3D 
would disappear (or be negligible for the desired simulation length). Efforts
to implement AMR are today, and will be for several years to come, central. \\

{\bf Multigrid Techniques}\\
When solving elliptic problems through standard relaxation schemes, it often is the case
that the low frequency modes of the solution (picturing the solution in
Fourier modes) are accurately obtained with relatively little computational effort while the higher modes
require substantially much more and are responsible for most of the computational cost. To alleviate this problem,
multigrid techniques\cite{thomasII} are introduced. The basic idea of multigrid is to eliminate the high
frequency 
components of the error quickly on a fine grid. These modes can be easily isolated by
transferring to a coarser grid and comparing the solutions. This strategy is carried out
through successive coarsening of grids and the results are transferred back to the fine grid.
The use of multigrid techniques has in the past been restricted to the initial 
value problem\cite{cookID,arnold,pedro,pedroID_II} and to solve the maximal slicing
condition in unconstrained implementations\cite{brugmanAMR}; but are now also being employed on
partially constrained evolutions in an axysimmetric code\cite{mattGRAXI}. \\

{\bf Parallelism} \\
Einstein equations are ideal candidates for constructing parallel implementations
which take advantage of supercomputers. The hyperbolic character of the equations
translate into the fact that to update the value of a field at a given point, only
a fine amount of information from the previous slice is needed. Hence, the computational
domain can be subdivided into smaller ones. Different processors/machines solve the equations
in these smaller cells and the solution is obtained at a later time after properly communicating
data among cells. This strategy
would imply that the elapsed time of a simulation $T$ on a single processor could in principle
be shortened to $T/n$ (if $n$ is the total number of processors used and neglecting the
overhead from the communications). In practice this is not
exactly the case but instead $T/(\alpha n)$ (with $\alpha<1$). Typical implementations
give $\alpha\in[0.7,1)$, which, although not `perfect' still implies that the more processors
used, the sooner the results will be obtained. Equally important is that the total memory
available is now $M_T = n M_1$. Hence not only can we obtain our solution sooner but we have
much more memory at our disposal to treat larger/more refined problems. 

As a last point, I would like to mention that until a very recent past, only very expensive
supercomputers provided researchers with enough computational power for
achieving large simulations. Unfortunately these supercomputers were not available
to all researchers. Their high cost and laws prohibiting the importation of such machines to many countries
prevented many from having access to powerful enough computers. Fortunately,
the picture is changing by the possibility of clustering many relatively low-cost machines
(like PC's) in what has been called `Beowulf supercomputers'. These machines will 
enable numerical relativists around the world to carry out their research more effectively
which will certainly have a positive impact on the field.\\

\subsubsection{Expediting the computational science aspect\\}
A particular aspect when exploiting the available computational power is the design of efficient codes.
Writing codes is very time consuming. The resulting product should not only minimize the amount of
computation and memory employed but also pay close attention to the way memory is 
being used (an efficient memory usage can speed up the performance considerably); input and output
is performed and the way data is to be stored. Taking care of these issues often exceeds the capability 
or the available time of numerical relativists who need to spend time concentrating 
on getting the physics correctly. It would be ideal if 
computer scientists could take care of the code's efficiency. Of course, having direct computer science
assistance is unlikely to be the case; but
fortunately, something is indeed being done in this direction. There exist software designed to expedite
writing efficient codes. Namely,
these software are capable of managing the memory usage, input/output, parallelization issues, 
data storage and helping in the implementation of AMR. 
Among these (freely-available) softwares products are RNPL\cite{marsa-phdthesis,rnplmanual}; 
PARAMESH\cite{PARAMESH}; PETSc\cite{PETSc}; KELP\cite{KELP} and the CACTUS Toolkit\cite{cactus}.  \\

{\bf RNPL}
lets the user simply specify the equations to be solved and how boundary conditions are to be treated 
and the compiler produces
the code. Remarkably, with little effort from the user, a code can be obtained.

{\bf PARAMESH} is a package of Fortran 90 subroutines designed to provide 
a relatively easy route to extend an existing serial code (which uses a logically cartesian structured mesh)
into a parallel code with adaptive mesh refinement (AMR).

{\bf PETSc} provides a suite of data structures and routines to write a parallel implementation
of a system governed by partial differential equations.

{\bf KELP} is a framework to implement parallel applications providing run time support
for blocked data decompositions. These block need not be uniform in size and AMR can be easily achieved
by appropriately chosen block sizes.

The {\bf CACTUS computational toolkit} was designed as a collaborative tool where users can adopt modules written by others
for specific purposes. In its bare bones, the users can choose to have the software handle the parallelization,
memory management and input/output and just concentrate on the physics per se.

As opposed to RNPL, all
other mentioned packages will
not write the code but provide an infrastructure which expedites the parallelization of the code,
incorporation of AMR and appropriate I/O and memory management.

Although these tools are not ideally suited for all problems; they can certainly
help researchers concentrate on the physical implementation without the need to spending 
too much time in the computer science aspect in a considerable number of situations.

\subsection{Analytical properties and numerical implications.}
The rich theory of PDE\cite{johns,geroch}, tell us a great deal of generic properties
of the expected solution. The distinction of hyperbolic, elliptic and parabolic teach
us how the system governs the way signals `propagate'; which data is needed to obtain a solution;
whether this solution exists and is unique; etc. I will here comment on two particularly
interesting issues regarding the interface between PDE theory and numerical 
implementations.

\subsubsection{Well posedness.}
Of particular importance is the concept of well posedness\cite{couranthilbert}. A well 
posed system is such that the solution $S$ (at time $t$) corresponding to the
initial data $u$ (at time $t=0$) can be bound by
\begin{equation}
||S|| \leq K e^{a t} ||u|| \, , \label{eq:wellposed}
\end{equation}
with $\{a,K\}$ constants independent of the initial data. (Note that this does not rule exponentially
growing solutions). Two can be cited as the main conclusions to be drawn from this property:
\begin{itemize}
\item{The growth of the solution is bounded. Although exponentially growing solutions are admitted,
there is an ``upper" limit to their growth rate.}
\item{The solution depends continuously on the initial data. }
\end{itemize}

In numerical implementations, clearly, the specified initial data in general will only be an approximation
to the desired initial data (since at best it can only be defined up to round-off errors); well posedness
guarantees (at the analytical level) that the obtained solution will nevertheless be in the neighborhood
of the solution we seek. Most systems being used in 3D Numerical Relativity are not known to be well posed,
(the exceptions being\cite{spectral2,bardeenPRIVATE,shinkaiASHTEKAR} in the $3+1$ approach, 
the conformal Einstein equations approach\cite{peterEVOL,frauendienerEVOL}
and the double null approach\cite{hpgn,bartnikALGOR}). 
The `danger' with systems that are not well posed is that $a$ in eq. (\ref{eq:wellposed}) might depend on
the initial data and therefore, the solution might have varying exponential growth
rates. In particular, it often is the case that if the initial data is ``pictured'' in terms of Fourier modes,
different frequencies $\omega$ have different values of $a$ and further
\begin{equation}
\lim_{\omega \rightarrow \infty} a \rightarrow \infty  \, .
\end{equation}
Note that, an unstable numerical implementation exhibits this behavior even if the system is well posed.
Of course, in practice $\omega$ does not attain infinity but, as the grid is refined, larger frequencies
are allowed and the solution {\it grows with the number of timesteps}! A behavior of this sort
has been investigated in the ADM system\cite{gomezADM,millerADM} for particular
gauge choices. The growth of $a$ with respect to
$\omega$ is not a `violent' one and, in principle could be controlled with the introduction of dissipation which
would keep the high frequencies in check. Further investigations will show if this is indeed the case.
I would expect that the discretization of a well posed system {\it should} simplify the attainment of
a stable numerical implementation, even though to date it has not yet clearly shown its advantages in this sense. 
As we learn more on how to exploit this feature, its role in the simulations will become increasingly useful.

A particular example from which conjectures can be drawn is the wave equation written in well posed
form and not.
Consider the following two systems obtained from $F_{,tt}=F_{,xx}$.\\
{\bf System (A)}
\begin{eqnarray}
F_{,t}=\Phi \, ,\\
\Phi_{,t} = G_{,r} \, , \\
G_{,t}=\Phi_{,r} \, ,
\end{eqnarray}
(where the intermediate variables $\Phi=F_{,t},G=F_{,r}$ have been introduced to reduce the original
system to first order).\\
{\bf System (B)}
\begin{eqnarray}
F_{,t}=\Phi \, , \\
\Phi_{,t} = F_{,rr} \, .
\end{eqnarray}
System (A) can be easily shown to be well posed while system (B) is not well posed in the usual sense. {\it Can we at least
say something on the expected behavior of the solutions of system (B), $S(B)$, given that we know how that from system (A) 
behaves?} Note that well posedness of (A) means that its solution $S(A)$
\begin{equation}
\fl ||S(A)|| = ||F(t)||+||G(t)||+||\Phi(t)|| \le \alpha e^{Kt} \big ( ||F(0)||+||G(0)||+||\Phi(0)|| \big ) \, ;
\end{equation}
since (at the analytical level) $||F(t)||+||\Phi(t)|| \le ||F(t)||+||G(t)||+||\Phi(t)||$ we can infer
\begin{equation}
\fl ||S(B)|| =||F(t)|| + ||\Phi(t)|| \le \alpha e^{Kt} \big ( ||F(0)||+||G(0)||+||\Phi(0)|| \big ) \, .
\end{equation}
Although this results does not imply well posedness (as the solution of system (B) is not bounded by its
initial data), it at least tell us that there is indeed an upper bound for the growth of the solutions.
This property could in principle address one of the criticism to symmetric hyperbolic formulations of Einstein equations, 
the large number of variables involved. One could start by considering one of these hyperbolic formulations
and then, replace the variables introduced to reduce the system to first order by the original higher
order derivatives. (Note that this `backtracking' can only be done if constraints were not added to the
`evolution' equations of the intermediate variables to achieve well-posedness). The obtained system would have considerably fewer variables 
and its solutions should still
be bounded. This approach is only recently receiving attention and it does appear to provide better 
behaved evolutions\cite{tiglio,lehnerDISS}. These preliminary investigations have been restricted
 to 1D, and further studies must be carried
out before firmer conclusions can be drawn.
At present, a clear advantage exploited from hyperbolic systems is the distinction of incoming variables
at a boundary (which are the only ones one is allowed to specify). \\

\subsubsection{Well posedness... is not enough!}
As mentioned previously, the difficulties observed in the numerical implementations of
the ADM equations lead to formulation of a number of symmetric hyperbolic systems. 
However, implementations of these systems did not show a significant improvement
in the obtained simulations. This is certainly not a surprise as well posedness
{\it does not} rule out the presence of exponentially growing modes. Moreover, in analysing
whether a system is symmetric/strongly hyperbolic one concentrates only on the principal part.
However, the non-principal part of the system can play a crucial role in the stability of
a numerical scheme. As an illustration, consider the following equation
\begin{equation}
f_{,t}=f_{,r} + f^2 \label{toyeqn} \, ,
\end{equation}
which is strictly hyperbolic and its principal part is just the 1D wave equation. There exists an
extensive set of algorithms capable of accurately treating the wave equation, however, the
addition of the $f^2$ term makes implementing equation (\ref{toyeqn}) delicate. In particular,
suppose one were to provide as initial data $f(t=0,r) = r^{-1}$ and boundary condition $f(t,r=R)=R^{-1}$.
The unique solution of such problem is $f(t,r)=r^{-1}$. Let's consider the linear perturbation
of (\ref{toyeqn}) in the neighborhood of this static solution. 
\begin{equation}
\delta f_{,t} = \delta f_{,r} + (2/r) \, \delta f \label{toyeqnII} \, .
\end{equation}
{\it What kind of solutions are allowed for such an equation?} Introducing the Fourier modes $\delta f = e^{su+ikr}$,
and replacing in (\ref{toyeqnII}) to solve for $s$, one obtains,
\begin{equation}
s=i \, k + (2/r) \, . 
\end{equation}
Thus, although the wave equation admits only purely imaginary values of $s=ik$,
our toy model, whose principal part is the wave equation does admit exponential modes. (Note that
if we had obtained a negative sign in front of $(2/r)$, we would have exponentially decaying modes
and, at least at the linearized level, the system would naturally drive towards the static solution).
In  the numerical realm,
one can readily see via the usual Von Neuman analysis\cite{kreissBOOK} that a straightforward extension
of stable schemes for the wave equation, lead to
unstable implementations of equation (\ref{toyeqn}).

The reader at this point might wonder why such particular example was chosen; after all, one
could always perversively modify an equation to display an exponential behavior. However, it turns out
that this simple example has a strong relationship with Einstein equations expressed in the 3+1 approach.
Recall equation (\ref{3plus1k}) for the evolution of the extrinsic curvature,
\begin{equation}
d_t K_{ij} = \alpha \left[
		{R}_{ij} - 2K_{i\ell}K^\ell_j + K K_{ij}\right] - D_i D_j\alpha \, .
\end{equation} 
It precisely has the form,
\begin{equation}
\partial_t K_{ij} = \beta^l \partial_l K_{ij} + f_1 (K_{ij})^2 + \mbox{extra terms} \, ;
\end{equation} 
with $i,j$ fixed and where $f_1$ is a function of the variables {\bf not including} $K_{ij}$.
If $f_1 > 0$ then an analogous local mode analysis indicates the presence of exponentially 
growing modes. {\it Is there anything one can do in this situation to `change the sign' of $f_1$?}
Note that we have at hand the constraints which can be arbitrarily added to the equations.
In particular, the Hamiltonian constraint has combinations of undifferentiated extrinsic curvature
components and, in principle, by adding it with appropriate factors one can `effectively' achieve 
the desired sign change or, the magnitude of $f_1$ be made much smaller.
An illustration of such procedure has been studied in the 1D case for the simulation of
Schwarszchild spacetime\cite{pabloK,lehnerDISS}. A remarkable improvement is obtained; without
the addition of the Hamiltonian constraint to the evolution of the extrinsic curvature, simulations
past $500M$ could not be achieved for all possible evolutions. With the modification of the
equations, stable configurations were obtained for all configurations. Note that although
the example presented here applies to the ADM formulation; all other 3+1 formulations have (at least some)
equations containing wave operators in the principal part and non-linear terms in the non-principal part where
a similar structure can be identified.

A related work has been presented in the 3D case with an implementation of a hyperbolic system
obtained by modifying the Einstein-Christoffel system\cite{yorkEINSTEINCRIST} by adding the constraints with free
parameters\cite{generalEC}. By simply varying the value of these parameters full 3D evolutions of 
single non-spinning black holes are achieved with evolution times ranging from a few $M$ to 
$1200M$. These results highlight the need for a deeper understanding on the influence of the non-principal
part of the system. Clearly, numerical implementations can considerably benefit from adding
the constraints in an appropriate way. Perhaps the simplest and quite general way of choosing
`ideal parameters' would be to do so by monitoring the evolution of the variables in a similar way as
artificial viscosity is often added in the numerical treatment of the hydrodynamic equations (see section \ref{matterdiss}).\\

\subsubsection{Elliptic equations and black holes.}
A recurring issue in numerical relativity is the role that constraints play in the evolution
of the equations. Analytically, they should be propagated by the evolution equations\cite{yorkID,frittelliCONSTRAINT}; 
numerically, as shown by Choptuik\cite{mattCONSISTENT} if the equations have been consistently implemented, the
constraints should be satisfied to the level of the implementation. These results justify the construction of
{\it free evolution} codes (ie. not dynamically enforcing the constraints as part of the evolution) and, in practice, 
the constraints
are monitored to show the quality of the obtained solution.
The use of free evolutions in black hole spacetimes (where singularity excision is to be used)
has also been preferred as it is not clear which boundary conditions are to be specified at the inner boundaries
(ie. those surrounding the excised singularities). Since the constraints are elliptic, the theory of PDE
tell us that the choice of boundary condition determines the solution globally (ie. there is an `infinite propagation
speed of signals'). This being the case the worry is that unless the correct data is known at the
inner boundary spurious solutions will result from a {\it constrained evolution}. In
fully or partially constrained systems\cite{mattMARSA,mattADSCOLLAPS}, inner boundary
conditions are obtained by employing the evolution equations to define values at the inner points for
all variables.  \\

\section{Particulars of Numerical Implementations of Einstein's equations}
In this section, I will very briefly review some aspects of the numerical
implementation of the formalisms described above.

\subsection{3+1 Approach:\label{evolcauchy}} 
{\bf Evolution Equations}\\
The evolution equations are implemented through, basically, the following structure
\begin{equation}
(\partial_t -{\cal L_{\beta}}) F = Rhs(F) \, ,
\end{equation}
where $F$ stands for the evolution variables and $Rhs(F)$ collects all extra terms. In order to treat this equation, 
the terms provided
by the Lie derivatives that include derivatives of $\beta$ are customarily moved to the right hand side.
\begin{equation}
(\partial_t - \beta^i \partial_i) F = Rhs(F) \, . \label{admevol}
\end{equation}
This splitting is carried out so that `standard' techniques developed for the
{\it advection} equation can be used to 
discretize this equation. The approaches most commonly used can be divided
in roughly two main groups: (i) {\it Operator Splitting} and (ii) {\it Straight discretization
of the right hand side}.
In the operator splitting strategy, the integration is divided into steps involving
parts of the original equation. One step integrates the homogeneous 
equation $(\partial_t - \beta^i \partial_i) F = 0$
while the other the `source' part $\partial_t F = Rhs$. Both steps can be intercalated, 
in different ways to produce an approximation for $F$ to a desired order.
For instance, 
\begin{eqnarray}
F^* = F^n + dt \beta^i \partial_i F \, ,\\
F^{n+1} = F^* + dt Rhs(F^*) \, .
\end{eqnarray}
This choice is by no means absolute, other options involve: integrating the source equation
first and then the homogeneous one and even proceeding in half steps\cite{MASSO}; treating
the transport part via-interpolations at the n-th or n+1-th level (examples of the algorithms used
are the cubic-interpolated pseudoparticle\cite{nakamuraCOLLNS}, 
causal-differencing\cite{seidelCAUSALDIFFERENCING,gundlachCAUSALDIFFERENCING,scheelCAUSALDIFFERENCING};
causal-reconnection\cite{miguelCAUSALRECONNECTION}, etc.). 

The second group involves a straightforward discretization of the right hand side. The most
promising approaches within FDA though, do make a difference in the way the $\beta^i \partial_i F$
term is treated (see for example~\cite{thornburgEVOL,peterEVOL,mattADSCOLLAPS,golmEXCISION}). These terms
are discretized using `up/down wind' type schemes where the sign of $\beta^i$ determines 
whether points to the right or left of the one under consideration are used. When using pseudo-spectral
methods, the right hand sides are evaluated straightforwardly and the method of lines is used to 
advance the solution to the next step\cite{spectral2}.\\

At present `3+1' unconstrained simulations are mainly based on a handful of formulations: the
 ADM\cite{mtw}; the BSSN (or `conformal
ADM')\cite{shibnakam,shapiroEVOL}, the `extended Einstein-Christoffel' formulation\cite{generalEC} and the
Bona-Masso formulation\cite{arbonaGAUGE}\footnote{Preliminary implementations
of Ashtekar formulation\cite{shinkaiASHTEKAR} have also been presented.}. The last two are
symmetric hyperbolic systems while the first two are not. The BSSN system is obtained from the ADM with the addition 
of extra variables
like the determinant of $\gamma_{ij}$; the trace of $K_{ij}$ and $\Gamma^k_{ij}$, coupled with a conformal
decomposition of the metric and extrinsic curvature and the use of the momentum constraint to replace some terms
in the resulting equations. The obtained system resembles the ADM one, but manages 
to (approximately) separate gauge dependent variables. When studying linearizations over flat space, the system
does indeed show appealing properties\cite{rendallITP,golmEVOL,millerADM}. Recently, several works
 have shown
the BSSN system provides longer evolutions than the ADM one. A peculiarity of the results displayed
by simulations obtained with this system is that the errors
in the constraints are larger than those obtained with the ADM one; nevertheless, as the evolutions proceed 
the ADM evolutions crashed earlier than those with the BSSN system\cite{shapiroEVOL,millerADM,golmEVOL}. The fact 
that the errors are larger could be explained by further
discretization errors introduced in the BSSN because of the extra variables evolved. These comparative studies
evolved both formulations with the {\it same} algorithms, however, there is no reason for the same `numerical recipe' to be
a good choice for both. Application of singularity
excision in the BSSN system has started recently, in $1D$\cite{lehnerCADM} it has shown similar results to those
obtained with the ADM one (for a specific way of handling the excision); recent $3D$ implementations show encouraging
results\cite{alcubierreSINGLEBH}.

The Einstein-Christoffel system implementation is presently being pursued using pseudo-spectral methods. Its hyperbolic
character has been exploited to simplify the treatment of both the inner and outer boundaries. 
Kidder et. al.\cite{spectral2} report
successful simulations of a single black hole in 1D (which is also `perturbed' via a Klein-Gordon field). The 
extension to 3D has been carried over with a related system (the extended Einstein-Christoffel system) achieving
evolutions of $1200M$\cite{generalEC} when constraint violating instabilities render the simulations inaccurate.\\

{\bf Inner Boundary}\\
As mentioned, when singularity excision techniques are used, an inner boundary appears in
the computational domain. This boundary is usually defined by finding the apparent horizon\cite{hawkell}.
In practice, to allow for displacements of the singularity a `buffer zone' is employed; ie. if 
the apparent horizon is located at $R=R(x^i)$, the inner boundary is placed at $R-\delta$ 
(with $\delta=n \Delta x$, $n \in[2,6]$). This buffer zone also allows the simulation to proceed
without needing to `locate' the apparent horizon at every timestep. Finding apparent horizons
is an `expensive' computational task. It involves solving an elliptic equation in 3D which 
defines a surface whose outgoing null normals neither diverge nor 
converge (ie. it is marginally trapped\cite{hawkell}). As usual with elliptic equations,
if a `good guess' is known, the task of solving it might not be so severe.
When a single apparent horizon is expected, a rough estimate of the mass of the hole coupled with
some notion of where the center of the horizon might be is exploited to yield fast apparent
horizon finders\cite{spectral3,huqAPPHOR,shibataAPPHOR,golmAPPHORIZ}. In the generic case, 
finding the apparent horizon can be a considerably expensive task, not only must the finder
be capable of starting with an arbitrary surface (usually chosen close to the boundaries of the
computational domain) and flow towards the location of the horizon, but also be capable
of handling several distinct apparent horizons. I am aware of only two of such
finders\cite{shoemakerAPHOR,dienerAPHOR} which are based in the {\it flow method approach} outlined 
in\cite{todAPHO}. To reiterate,  although finding apparent horizons
on a given surface is an expensive computational task, it need not be found at every single timestep.

{\it What is done at the inner boundary points?} As mentioned, the strategy is to use the 
evolution equations to update these points.
Both strategies employed at the `bulk' (which we mentioned in the previous point) are suited to
implement this idea. An important requirement is that
the shift is conveniently chosen in the neighborhood of the excised region. Namely, $\beta^i$
has such that $(\partial_t - \beta^i \partial_i) F = 0$ describes signals
propagating {\it towards} the excision boundary and not {\it from} it. If this were not the case, then,
it will be difficult to prevent signals propagating from regions inside the event horizon
to the outside. Although these methods appear to work reasonably well in lower dimensions, their
3D implementations are not yet robust enough (but considerable progress has been achieved in the past
year with single black hole evolutions being carried out for times beyond $500M$\cite{alcubierreSINGLEBH,generalEC})

The goal pursued by all methods is to have an accurate and stable implementation
of the Equations at the inner boundary (often called `excision boundary'). Note that when
using finite difference techniques the right hand side of the equations can not
be evaluated in centered way 
(as there are `no points' available at the interior of the excision boundary). In practice,
interpolation or extrapolation is used; this process must be handled 
with care as it not only introduces `high-frequency' features in the solution but in can also
render the evolution unstable\cite{lehnerCADM}. There are a number of methods under use, differing
in the way the interpolation is carried out and which of the previously mentioned groups (operator
splitting or straightforward discretization of the rhs) is adopted.
The techniques presently used
are:  causal-differencing\cite{seidelCAUSALDIFFERENCING,gundlachCAUSALDIFFERENCING,scheelCAUSALDIFFERENCING}
and more simple minded excision techniques with up/down wind algorithms\cite{golmEXCISION,thornburgEVOL}. 
With spectral methods, on the other hand, as one counts
with a continuous representation, the evaluation of the desired variable and its derivatives
can be made at any point without needing to interpolate. Hence, the right hand sides of the equations
are straightforwardly evaluated and the method of lines is used to advance the solution to the 
next hypersurface\cite{spectral2,generalEC}. 
It is important to point out that handling a moving singularity is a crucial test for a robust
treatment of the inner boundary (as points will `pop out' from the excision region and the 
evaluation of the eqns will shift location at different hypersurfaces). Only causal differencing
has been shown to be partially successful in this problem. One might argue that with appropriate coordinate
conditions, one can `fix' the singularity in the grid and therefore need not pass such a test.
However, it is difficult to imagine that such coordinate conditions will be available for all problems
and even if this is the case, treating a moving singularity will likely the limitations of the implementation.\\

A possible way to `aid' the numerical implementation
is to `modify' the equations near the excision region. Since, in principle, nothing can escape
from the event horizon, one could use this fact to simplify the implementation of
the evolution equations; for instance, consider the following variation of eq. (\ref{admevol})
\begin{equation}
\partial_t F - ( W \beta^i + (1-W) V^i)  \partial_i F  = W Rhs(F) \, ;
\end{equation}
where $W=1$ outside the apparent horizons and smoothly going to zero at the excision
boundary. The vector $V^i$ could be chosen appropriately so that signals propagate
normal to the excision boundary; be zero so that the values of the variables are frozen; etc.
I am aware of the use of an analogous strategy only in the implementations of the conformal
Einstein equations\cite{peterEVOL}, except that in this case was used to control signals from
propagating into the physical spacetime crossing ${\cal I}^+$.\\

{\bf Initial Data}\\
As mentioned, initial data must satisfy four constraint equations. For spacetimes free of singularities,
these initial data together with appropriate outer boundary conditions determine a unique solution\cite{york-sources}.

When singularities are present, either inner boundary conditions are prescribed or, if possible, 
the singular behavior removed from the field variables. In the past, most efforts towards obtaining
valid initial data were carried out under certain assumptions which, although restrictive, considerably
simplified the treatment and allowed gaining valuable experience in treating this problem (see for
instance \cite{cookID,puncturesID}).
For instance, the families of Brill-Lindquist\cite{brilllindq} and Misner\cite{misner} data provide
multi-black hole solutions under the assumptions of conformal flatness and time-symmetry.
Relaxing the time-symmetric assumption, but still keeping conformal flatness, provides
more generic multi-black hole solutions referred to
 as `Bowen-York'\cite{bowenyorkID} data and `puncture' data\cite{puncturesID}.

These data sets have several drawbacks for astrophysically relevant applications. 
One is the assumption of conformal flatness, as has recently been shown by Garat 
and Price\cite{priceCONFORMALFLAT}, there exists
no spatial conformally flat slicings for the Kerr spacetime. Therefore, even in a spacetime containing a
single spinning black hole, the assumption of conformal flatness introduces unphysical radiation. Further questions 
on the suitability
of the Bowen-York solutions for astrophysically relevant simulations have been raised in\cite{loustoID}. By
considering the `particle-limit' of these data sets, the authors find that even in the case of a single
non-spinning black hole spurious radiation is present. Additionally, 
these solutions are all obtained on a maximal slice (in the case $\gamma^{ij}K_{ij}=0$) which allows for the constraint
equations to decouple but considerably restricts the available freedom

The aforementioned initial data sets have proved quite valuable in investigating different aspects of
the theory and  numerical implementations of black hole spacetimes. As the focus turns to producing
astrophysically useful information, a revision of the initial data specification is required.  
Recently a number of proposals have been introduced where conformal flatness has
been dropped\cite{jeffID,matznerID,pedroID_II,dainID}. Here, the Lichnerowicz-York approach is still used,
the difference lies in the non-flat `seed' metric $\hat \gamma_{ij}$ provided. As a result, the constraint
equations are coupled and must be solved simultaneously. The approach introduced in\cite{matznerID,pedroID_I},
has recently been fully implemented in 3D where $\tilde \gamma_{ij}$ has been chosen to be the superposition of
boosted Kerr black holes\cite{pedroID_I}. By conveniently `weighting' this superposition, reasonable inner boundary data
around each (excised) singularity can be induced from the analytically known single black hole solution. \\

\subsubsection{Examples of implementations\\}
{\bf 1D}\\
Spherically symmetric spacetimes still offer a rich arena to study strong gravity effects.
Applications in critical phenomena, collapse simulations, singularity structure studies, etc. are
within reach of reliable simulations. Additionally, 1D simulations are useful first steps to test algorithms
for more generic spacetimes. \\

{\bf 2D}\\
Simulations assuming axysimmetric spacetimes are being carried out to investigate
critical phenomena\cite{mattADSCOLLAPS,garfinkleCOMPACT}, black hole collapse situations,
rapidly rotating neutron stars\cite{shibataROTATINGNS}, black hole accretion physics, etc.
Here, the problem of the coordinate singularity at the symmetry axis must be addressed.
This is done by enforcing regularity conditions at the axis\cite{bardeen-piran} or by 
``thickening'' the
direction along the spacelike killing vector so that enough points are available to take
derivatives as if it were a 3D spacetime (and then interpolate the results back to define their values
at the axis)\cite{alcubierre2D}. Preliminary investigations of gravitational wave collapse 
scenarios\cite{garfinkleCOMPACT} 
display critical behavior of the solution; these simulations are still rather coarse and
more definitive results will be obtained with the use of AMR.\\

{\bf 3D}\\
3D simulations are mainly targeting black hole/neutron star systems. Studies of collapse of
compact objects or collapse of waves onto black hole are being pursued. 
Considerable
progress has been obtained as the first series of simulations are being
reported\cite{nakamuraCOLLNS,suenNSCOLL,shibataCOLLNS,usCOLL,golmCOLL2,alcubierreSINGLEBH}.
As discussed, 3D Numerical Relativity
is very challenging already from the computational-resources point of view;  this has
restricted the resolution used in all these works.
All the obtained models have been able to simulate the systems under study for moderate
amount of times, enabling preliminary conclusions to be drawn from them. The focus is now to address
the observed stability problems and improve the resolutions. Perhaps
many of the stability problems faced so far 
might disappear, or become negligible for the targeted simulation length, when fine enough resolutions
can be achieved.

\subsection{Characteristic\label{causticfix}.} 
{\bf Evolution equations}\\
The evolution equations in this formulation are implemented observing 
that the left hand side of the equations correspond, roughly, to wave 
equations in $(u,r)$ coordinates, ie.
\begin{equation}
2 (r h_{AB} )_{,ur} - ( (V/r) (r {h_{AB}})_{,r} )_{,r} = RHS \, . \label{evolcharac}
\end{equation}
A crucial ingredient is the way tensor fields (and derivatives) appearing in the
RHS (of the evolution and hypersurface equations) are handled on
the spheres $r=const, u=const$ and that a single patch can not be used to cover these
sphere. Efficient implementations have been obtained with the 
use of {\it eth}-operators\cite{goldberETH}, which have been implemented 
via second order FDA\cite{eth} or through the use of Fast-Fourier 
transformations\cite{bartnikALGOR}. Recall that inner boundary conditions are required;
once these have been specified, integration of the hypersurface equations is
carried out by explicit second order FDA\cite{cce,dinvernoEVOL},
or by and 8th order Runge Kutta integration\cite{bartnikALGOR} marching radially outwards.
Finally the evolution equations are integrated explicitly in time and no outer boundary
conditions are required as the last point on radial lines lies on an incoming null
surface ${\cal I}^+$. \\

{\bf Caustics}\\
The common disadvantage of all characteristic codes is the necessity to
either deal with caustics or to avoid them. It has been proposed
to treat these caustics ``head-on'' as part of the dynamical problem\cite{friedrich-stewart}.
Since only a few structural stable caustics can arise, their geometrical properties
are well understood and their behavior could be treated numerically\cite{stewart-friedrich}.
To date, this option has not been pursued but its beauty and potential can not be denied.
In the mean time, the formulation can be used in:
\begin{itemize}
\item{Spacetimes where caustics will not render the coordinates singular. For instance when
dealing with compact objects, the lens equation provides a rough estimate of
when they can appear\cite{nigelNEW}.}
\item{Spacetime {\bf regions} without caustics. Here, the use of Cauchy-characteristic
matching (CcM)\cite{bishopccm,dinvernoCCM,matching,jeffccm} exploits the main advantages offered by ``3+1''
and characteristic codes. A ``3+1'' formulation is employed to simulate strong curvature regions in
a bounded domain, on the exterior (which is assumed free of caustics) 
of that domain a characteristic formulation is employed. The combination
manages to cover the entire spacetime, removing the boundary problem for the ``3+1'' code and, the caustic
problem for the characteristic one. Although CcM is not yet satisfactorily working in 3D, its successful applications
in simpler cases illustrates its usefulness (see for instance\cite{Ccpaper,dinvernoCCM})}.
\item{Combination of regions patched with different characteristic codes. Characteristic-characteristic
matching (c2M)\cite{c2m}, can also be used to avoid caustics while simulating the whole spacetime (although
can be used in a more restrictive set of problems than CcM, its implementation in 3D should be rather 
straightforward\cite{c2m}).}\\
\end{itemize}

{\bf Initial Data}\\
As discussed, another distinctive feature of a characteristic formulation is that
the initial data is constraint-free. Namely the intrinsic (conformal) metric $h_{AB}$
is freely specifiable on an initial hypersurface ${\cal N}_0$ and the integration of the
hypersurface equations (which are basically ODE's) provide the complete metric
on ${\cal N}_0$. This trivializes posing consistent initial data; however, the problem
of defining data which conforms to the physical situation in mind still remains.
For the vacuum case, a convenient option is to set the Weyl component $\Psi_0=0$ (in the
language of the NP formalism\cite{NP}), this choice minimizes the radiation crossing
${\cal N}_0$ when the departure from spherical symmetry is small\footnote{Note  that if the initial
 null hypersurface coincides with
${\cal I}^-$ this is precisely the condition of no incoming radiation}. For the case of spacetimes 
with non-trivial matter content, a consistent way of defining the intrinsic metric was introduced 
by Winicour\cite{jeffnewtonian}. Contact with post-Newtonian theory is obtained through
a perturbative analysis with a varying speed of light. The obtained prescription is such that
the radiation observed at ${\cal I}^+$ reduces, to first order, to the 
familiar quadrupole approximation.

\subsubsection{Examples of implementations\\}
{\bf 1D}\\
There is a considerable wealth of 1D characteristic codes which have been applied to
study: the radiation tail decay of spacetimes containing scalar fields\cite{gomezSCALAR};
critical phenomena\cite{hamadeCRITICAL,garfinkleCFINCHARACT,charac_crit}; singularity
structure\cite{gnedinCH,bradyCH,burkoCH,piranCH}; scalar fields as precursors of inflationary 
cosmology\cite{maddenCHARACT}; cosmic strings (represented by massive scalar
and vector fields coupled to gravity)\cite{sperhakeCOSMICSTRING} and self-similar
collapse of spherical matter and charge distributions\cite{barretoMATTER}, among others.\\

{\bf 2D}\\
A 2D characteristic code for twist-free axisymmetric vacuum spacetimes was
developed in\cite{papa2D} and recently been extended to handle matter
through the use of high resolution shock capturing schemes\cite{siebelNS2D}. This 
implementation is being applied to study neutron stars in full GR. Another implementation that 
removes the twist-free requirement has been presented\cite{bishopdinverno,dinvernoCCM} and is
being employed in a larger Cauchy-characteristic matching code 
(the Cauchy code used is the axisymmetric ADM code introduced in\cite{piran-stark}. 
A double null code (under the assumption that departures from spherical symmetry
are small) has been employed to simulate a region exterior to the event horizon of the Kerr-Newman spacetime.
The inner boundary is placed at the incoming null surface defined by $r=3m$ (with $m$ the mass
of the black hole)\cite{shinkaihayward}.
Another recent implementation\cite{husainCHARACT} has been used to study scalar
field collapse in spacetimes with negative cosmological constant. Aside from the
study of black hole formation, the interest in anti-deSitter spacetimes from 
AdS/CFT proposed duality in string theory makes this an important subject\cite{REFDUALITY}. Although the 
conjectured duality between AdS spacetimes and physical effects in conformally
invariant Yang-Mills theories on its boundary is for five dimensional spacetimes, 
the work presented in\cite{husainCHARACT} appears as a natural first step for numerical studies
of this duality.\\ 

{\bf 3D}\\
There exists two characteristic codes in 3D. The first one, obtained by second order
accurate FDA has been presented in\cite{cce,hpgn} for the vacuum case was and used to 
simulate black hole spacetimes (for `unlimited times' $\approx 60000M$ with $M$ the mass of the 
black hole) and study scattering off 
Schwarzschild black hole in the highly nonlinear regime (stably simulating power outputs up to $10^{60}W$). 
Notably,
the transition from 1D to 3D is considerably simplified by replacing tensors by spin-weighted
complex scalar fields and angular derivatives by eth-operators\cite{goldberETH} (which are in turn
implemented by FDA and interpolations between the two patches used to cover spheres at $r=const$\cite{eth}).
At present this 3D code is being extended in two directions. On the one hand, the equations governing a perfect
fluid have been incorporated (in a rather crude way) for a feasibility study of simulations
of black hole spacetimes containing a companion star. Encouraging results were
obtained in collapse of dust or matter with weak pressure onto a black hole\cite{mattchar}. At present, more 
realistic matter data is being studied and plans for incorporating high resolution shock capturing
schemes\cite{philipMATTER} are under way. On the other hand, a project aimed towards
obtaining gravitational radiation of a binary black hole spacetime is also under development\cite{jeffPROGRESS}.
Here, the spacetime is envisioned in a time-reversed point of view. This is motivated by
the possibility of posing a double null problem whose inner boundary corresponds to
a fissioning white hole\cite{fissionwh,husaTORUS} (which in a time reversed point of view corresponds
to merging black holes)
and the other corresponds to ${\cal I}^-$. An inverse scattering process can be formulated
to obtain the radiation produced by a binary black hole collision\cite{gomezFISSION}. Preliminary investigations
 of this approach
have targeted a ``close limit approximation'' yielding excellent results\cite{jeffFISSIONCLOSE}.

In an independent 3D implementation\cite{bartnikALGOR}, a characteristic code has been developed
not in Bondi-Sachs coordinates but rather using a null-quasispherical gauge\cite{bartnikFORM}. In this
gauge, the angular part of the metric is effectively a unit sphere metric (this can always be done as
surfaces at $u=const, r=const$ have $S^2$ topology). The angular coordinates transformation (which naturally
depends on time), encodes the radiation content of the spacetime. The numerical implementation is obtained
through (I) a clever combination of FDA, fast-Fourier transforms and spectral decomposition of tensors in terms
of spin-weighted spherical harmonics to handle fields on the spheres; (II) an 8th order Runge-Kutta integrator
for the hypersurface equations and (III) the method of lines with a 4th order Runge-Kutta time stepper. 
This code has been used to study (linear to mildly non-linear) scattering off a (mass $M$) Schwarzschild Black Hole.
The resulting simulations exhibit very high accuracy and evolutions for about $~100M$ are reported, the evolution
terminates at late times close to the event horizon where the null-quasispherical gauge apparently breaks down.\\

\subsection{Conformal}
{\bf Evolution Equations}\\
The evolution equations formally look very much like those discussed in section~\ref{evolcauchy}.
Codes implementing the  conformal evolution equations have been obtained using standard
FDA for both the time and spatial derivatives\cite{frauendienerEVOL} (in 2D) or
have employed the method of lines\cite{peterEVOL} (in 3D), where FDA approximations are used
for the spatial derivatives while the time integration is carried over by a standard
4th order Runge-Kutta algorithm.\\

{\bf Outer Boundary}\\
Specifying boundary values for the evolution part is simplified in this formulation
as it need not conform to the physical problem in mind. This might appear puzzling at first
sight but let's not forget the outer boundary
is causally disconnected from the physical spacetime; hence, 
in principle one can pose arbitrary conditions as long as this is done
in a stable manner. Furthermore, even the equations might
be modified in the unphysical region to aid in this task.  In\cite{peterEVOL}, the evolution
equations are modified (beyond ${\cal I}^+$) to mimic advection equations describing signals
 propagating towards
the outer boundary and therefore ``numerical diffusion'' which could leak into the 
physical spacetime is minimized.\\

{\bf Initial Data}\\
Initial data is obtained by solving the Yamabe equation (obtained from the
Hamiltonian constraint)\cite{frauendienerreview} 
 in such a way that its degeneracy at the boundary is properly addressed.
Pseudo-spectral methods are employed which aid in obtaining solutions with the 
proper regularity conditions\cite{frauendienerID,peterID}. Data corresponding to
flat spacetime, vacuum spacetime with toroidal infinities\cite{A3} and 
Schwarzschild spacetime\cite{schmidtSCHW}, among
others, are available.\\

\subsubsection{Examples of implementations\\}
{\bf 1D}\\
Scalar field collapse situations were studied by Huebner in\cite{peter1D}, reproducing
the scaling law behavior obtained by Choptuik\cite{mattFIRSTCRITICAL}, but in this case, being able
to simulate the full spacetime.\\

{\bf 2D}\\
Frauendiener\cite{frauendienerEVOL} implemented a 2D code to 
study A3-like space-times\cite{A3}. These provide the first examples
of vacuum space-times with gravitational radiation. Although the toroidal topology
of future null infinity imply they cannot be used as models of isolated systems,
they provided a rich arena to investigate the system and calibrate the implementation
in higher dimensions.\\

{\bf 3D}\\
Quite recently, a 3D implementation was used to simulate
the Schwarzschild space-time\cite{peterITP}. In particular, the full Kruskal
diagram was targeted and encouraging results were obtained as a significant
portion was accurately simulated. Additionally, the code has been used to study
initial data sets departing slightly from flat spacetime\cite{peterLAST}. The simulation is able to
reproduce the rigorous analytical results from Friedrich\cite{friedrichREGULAR} (and
related to those of Christodoulou and Klainerman\cite{chrisklain}) that these initial data should evolve in such
a way that a regular $i^+$ should exist. The entire future of the initial hypersurface
is accurately obtained and the radiation at ${\cal I}^+$ is extracted; to date this is
the most complete simulation of this kind of system.\\

\section{Beyond the Vacuum case \label{matter}}
\subsection{Scalar Field Models}
Although scalar fields have not been observed in nature so far, their study 
has been carried out since the 60's\cite{kaup,ruffini}. The original motivation
for them was to consider the existence of bosonic counterparts of
observed fermionic objects (like neutron stars). These objects can provide useful
physical insights in a variety of fronts since they are sources of scalar
gravitational radiation and can collapse to black holes. More recently, these objects
have been suggested as candidates for dark-matter\cite{madsen} thus being ``promoted'' from purely
theoretical toy models to perhaps real physical objects.
An important feature of the scalar field models under study is that they do not
develop shocks or discontinuities (if these were not already present in the initial data) 
which simplifies their numerical simulation. Not only have scalar field models been useful
to investigate: ``stability'' of Minkowski spacetime; critical phenomena; singularity structure;
cosmological models; alternative theories of GR; etc., but also have served well to test codes for
their use in relativistic hydrodynamics.

A large number of scalar field models exist, these have been introduced considering
both real and complex fields which can be massive and/or charged. For simplicity I 
will next consider a simple case, that of the massive Einstein-Klein-Gordon field\cite{wald}
to illustrate their use. The real scalar field $\Phi$, satisfies the equation
\begin{equation}
\nabla^a \nabla_a \Phi = m^2 \Phi \, \label{scalareqn};
\end{equation}
which is derived by minimizing the action
\begin{equation}
S = \int [ R - (\frac{1}{2} \nabla_a \Phi \nabla^a \Phi + m^2 \Phi^2) ] dV \, ;
\end{equation}
with $R$ the Ricci scalar and $m$ the mass of the field. The stress energy
tensor $T_{ab}$ is given by
\begin{equation}
T_{ab} = \nabla_a \Phi \nabla_b \Phi - \frac{1}{2} g_{ab}( \nabla_c \Phi \nabla^c \Phi + m^2 \Phi^2) \, .
\end{equation}
The dynamics of the scalar field is governed basically by a  wave equation in a curved
spacetime (\ref{scalareqn}). Particularly interesting is the possibility of 
stable (or long lived) compact configurations of complex massive scalar fields known as
{\it boson stars}. These are local equilibrium solutions of the system in which the spacetime
is static (although the real and imaginary components of the field oscillate). These `stars'
are `similar' to neutron stars in the sense of having a maximum mass marking a transition
from stable to unstable states. Additionally there exists a family of
solutions known as {\it multi-scalar stars}
which are quasi-periodic compact solutions to the Einstein-Klein-Gordon systems. This class
of solutions contains boson stars and {\it oscillating soliton stars} (periodic solutions
of systems with a single real scalar field). The study of boson stars in fully General Relativistic
scenarios was started by Seidel and Suen\cite{seidelBOSON,seidelBOSONII} to investigate their
role as a possible source of dark-matter. Since then, numerical simulations have been directed
towards analyzing stability of boson stars and critical phenomena\cite{schunckBOSON,hawleyBOSON}; 
investigate possible `boson halos'
around galaxies and their influence on them\cite{balakrishnaHALO} and simulate the collision of
`boson stars'\cite{balakrishnaTHESIS}.

\subsection{Relativistic Hydrodynamics}
In the non-vacuum case a fluid is characterized by its velocity $u^a$, pressure $p$,
enthalpy $\epsilon$ and rest mass density $\rho$ defined in a locally inertial reference frame.
The general relativistic hydrodynamic equations consist
of the local conservation of $T_{ab}$ (a direct consequence of the Bianchi identities)
and of the current density $J^a=\rho u^a$ (the continuity equation).
\begin{eqnarray}
\nabla_a T^{ab}=0 \, ; \label{fluid1} \\
\nabla_a J^a = 0 \, . \label{fluid2}
\end{eqnarray}
These equations determine the dynamics of the fluid, while 
Einstein equations (appropriately modified to include the corresponding
components of $T_{ab}$ on the right hand sides) determines the geometry.  When
neglecting non-adiabatic effects (such as viscosity or heat transfer) the stress energy
tensor for a perfect fluid is,
\begin{equation}
T_{ab}=\rho h u_a u_b + p g_{ab} \, ;
\end{equation}
with $h$ the relativistic specific enthalpy given by $h=1+\epsilon+p/\rho$. In order for
the system be solvable, the five equations (\ref{fluid1},\ref{fluid2}) must be supplemented
with {\it two} extra conditions. One of these is $u^a u_a=-1$ and the other an
equation of state $p=p(\rho,\epsilon)$.

An accurate simulation of this system is a challenging task even in Newtonian gravity. The
difficulty lies in the fact that the system develop shocks, rarefraction waves and contact
discontinuities which are difficult to handle (which, because of the non-linear character
of the equations governing the fluid, can develop even though they were not 
present in the initial data). To simplify the treatment of the system,  
equations (\ref{fluid1},\ref{fluid2}) are rewritten in explicit conservation form. This requires
introducing intermediate variables which are integrated on time, and the primitive variables
are recovered at each step by an, often expensive, inversion method. Flux conservative systems are formally
simpler to handle and simplify implementations where variable grid spacing is employed.

Most ways of expressing the equations were obtained for the 3+1 approach (namely the ADM one). Recently,
interest in covariant expressions which could be applied in different approaches resulted
in a number of re-formulations\cite{eulderinkCOVARIANT,philipMATTER}.

In\cite{philipMATTER}, the spatial components of the four velocity $u^i$ together with $\rho$ and $\epsilon$
are taken as primitive variables. The intermediate variables are $V^A=(\rho u^0,\rho h u^0 u^i + p g^{0i},\rho u^0 u^0
+p g^{00})$, $(A=0,i,4)$. In terms of $V^A$, the equations take the form
\begin{equation}
\partial_o (\sqrt{-g} V^A) + \partial_j (\sqrt{-g} F^j) = S \, ,
\end{equation}
with
\begin{eqnarray}
\fl F^j=(J^j,T^{ji},T^{j0})=(\rho u^j,\rho h u^i u^j+pg^{ij},\rho h u^0 u^j + p g^{0j}) \, , \\
\fl S^A=(0,-\sqrt{-g} \Gamma^i_{ab} T^{ab}, \sqrt{-g} \Gamma^0_{ab} T^{ab} ) \, .
\end{eqnarray}

After integrating these equations, the value of the primitive variables are recovered typically
by a root-finding algorithm like the Newton-Rapson one\cite{recipes}. This feature is computationally expensive
and might even lead to accuracy loses. However, in the case where a characteristic formulation
is employed, $g^{00} =0$ which allows for an explicit recovery of the primitive variables\cite{philipMATTER}.

\subsubsection{FDA and relativistic hydrodynamics.\label{matterdiss}} 
As mentioned in section~\ref{muchado}, FDA algorithms are obtained by formal Taylor expansions, this
naturally carries the implicit assumption that the variables are smooth enough for such expansion to be valid.
Clearly, discontinuities do not satisfy this requirement and in practice are ``smoothed-out''
via the addition of artificial viscosity terms to the stress energy tensor in the following
way
\begin{equation}
T_{ab} \rightarrow T_{ab} + Q_1 u_a u_b + Q_2 g_{ab} \, ,
\end{equation}
with $Q_1$, $Q_2$ `viscosity controlling functions' which can be chosen independently. For instance, in Wilson's formulation\cite{wilson}
$Q_1\equiv 0$ while in the one by Norman and Winkler~\cite{normanwinkler} both $Q$'s are allowed to be non-zero.
These extra terms are such that, as the grid is refined, they tend to zero (and therefore one 
does have a consistent approximation to the original system). In order to avoid dissipation in regions where
the solution is smooth, $Q$'s are defined to be non-zero only in places where the solution has large gradients.

Clearly, the magnitude of these terms must be carefully chosen so that the necessary amount of dissipation is introduced
but, at the same time, excessive smearing of the discontinuities is avoided. Assuming this can be done,
artificial viscosity is indeed very appealing as it is straightforward to
implement and computationally efficient. For these reasons, this technique enjoyed an absolute popularity for
more than three decades. It has only been until recently that other options, the {\it high resolution shock capturing
schemes}~\cite{martiHRSC}, have become popular. These methods exploit the hyperbolic character of the equations and 
explicitly use
the characteristic speeds and directions to solve (exactly or approximately) the Riemann problem at every
interface of the numerical grid\cite{fontreview}. This property guarantees
that physical discontinuities are treated consistently, producing stable and sharp discrete shock profiles while
providing good accuracy order. To illustrate the spirit of this technique, let's take the $1D$ case and define
$\Omega=\{(x,t), t\in [t,t+\Delta t], x\in[x_o,x_o+\Delta x]\}$; consider,
\begin{equation}
\partial_o (\sqrt{\gamma} V) + \partial_x (\sqrt{-g} F) = S \, ;
\end{equation}
can be formally integrated as
\begin{equation}
\fl (\bar U \Delta)|_{t+\Delta t} - (\bar U \Delta)|_{t}=
-\left( \int_{L_1}  (\sqrt{-g} \hat F) dt - \int_{L_2}  (\sqrt{-g} \hat F) dt \right )
+ \int S dt dx \, ;
\end{equation}
with $L_1=(x_o,t), L_2=(x_o+\Delta x,t)$ ($t\in[t_o,t_o+\Delta t]$)
\begin{eqnarray}
\bar U = \frac{1}{\Delta V} \int_{\delta V} (\sqrt{\gamma} U) dx \, , \\ 
\Delta V = \int_{x_o}^{x_o+\Delta x} \sqrt{\gamma} dx \, .
\end{eqnarray}
where $\hat F$ are the fluxes across the numerical cells which depend on the solution
at the interfaces. At them, the flow conditions can be discontinuous and can be obtained,
as Godunov suggested~\cite{godunov} by solving a collection of {\it local} Riemann problems.
In practice, the continuous solution is locally averaged on the numerical grid
leading to discontinuities at cell interfaces.  Accurate knowledge of the Riemann problem's problem
is exploited to obtain the solution at the later time. Dissipation is still added in the process 
but the information of the local characteristic of the fluid is used to do so in the ``correct'' amount.

\subsection{Other options}
Two approaches have been considered which can be regarded as hybrid combinations
of FDA for the geometric variables and a ``particle'' approximation for the 
fluid variables. These approaches are known as: {\it Smooth Particle Hydrodynamics} and
{\it Particle Mesh}.\\

\subsubsection{Smooth Particle Hydrodynamics}
In the smooth particle hydrodynamics (SPH) method, the fluid is modeled as a collection of
particles which are represented by smoothed values. That is, given a function $f(x^i)$ its mean
smoothed value $<f(x^i)>$ is obtained from
\begin{equation}
<f(x^i)>\equiv \int W(x^i,\hat x^i;h) f(\hat x^i)\sqrt{\gamma} d^3\hat x^i \, ;
\end{equation}
where $W(x^i,\hat x^i;h)$ is the kernel and $h$ a smoothing length. The kernel satisfies
\begin{equation}
\int W(x^i,\hat x^i;h)\sqrt{\gamma} d^3\hat x^i  = 1 \, ;
\end{equation}
gradients and divergences are also represented by smoothed counterparts; for instance,
\begin{equation}
<\nabla f(x^i)>\equiv \int W(x^i,\hat x^i;h) \nabla f(\hat x^i)\sqrt{\gamma} d^3\hat x^i \, .
\end{equation}
After introducing the density distribution of particles,
\begin{equation}
<n(x^i)>=\Sigma_{a=1}^N \frac{\delta(x^i-x_a^i)}{\sqrt{\gamma}}
\end{equation}
with $\{x_a^i\}_{a=1..N}$ (the collection of N-particles where the functions are known).
These approximations are used to derive a smoothed version of the general relativistic
hydrodynamics equations (\ref{fluid1},\ref{fluid2}). The explicit formulae are reported in\cite{pablo1}.
Again, viscosity terms must be introduced to deal with simulations where shock waves arise\cite{mann}.
The integration of the hydrodynamic equations via this method reveals only pair-wise particle
interactions among particles inside the compact support of the kernel. The drawback is the need
to search among all $N$ particles those $N_h$ in a given kernel. The use of hierarchical grid methods\cite{hierartree}
makes the search be an $O(N \ln N)$ task, once the search is performed, the update takes only $O(N_h N)$.
Studies of tidal disruptions by supermassive black hole spacetimes have been presented in\cite{pablo1,pablo2}
where the background is kept fixed. I am not aware of SPH being used to study a fully relativistic problem yet.

\subsubsection{Particle Mesh.}
In this approach, the fluid is treated as a ``collisionless gas of particles''.
The stress energy tensor is expressed as
\begin{equation}
T^{ab} = \Sigma_A m_A n_A u_A^a u_A^b \, ,
\end{equation}
where $m_A, n_A, u_A^a$ are the rest mass of the particle, the
number density in the comoving frame and the 4-velocity of each
particle. Each particle's evolution is determined by the geodesic
equation. The integration of the geometric variables using FDA requires
an interpolation of the stress energy tensor onto the grid points. Additionally,
the evolution of the particles requires interpolating the metric variables
onto the particle's trajectory. This method has been extensively applied
by Shapiro and Teukolsky to investigate stellar dynamics\cite{shapteukstellar},
collapse of dense star clusters to supermassive black holes\cite{shapteukcluster}
and the formation of naked singularities\cite{shapiroteukolskyNAKEDSPINDLE}.\\

\subsection{Initial Value problem}
Most works dealing with non-vacuum spacetimes and targeting astrophysically relevant simulations
employ 3+1 formulations\footnote{The exception being \cite{philipMATTER,philipACCRETION,nigelNEW} which
adopt a characteristic formulation.}. I will next comment on how initial data for these simulations is
obtained. 

In the non-vacuum case, the Hamiltonian and momentum constraints must be solved taking into
account the corresponding terms of (the now non-vanishing) stress energy tensor. From the
implementational point of view, little changes. Given appropriate definitions for the matter fields
$(\rho,p(\rho,\epsilon),\epsilon,u^a)$ the same modules used for the vacuum case can be used to obtain
the gravitational data. However, one is usually interested in situations where both matter and geometry are
in (or close to) equilibrium. That is, the spacetime is assumed to (approximately) have a timelike  
killing vector.

\subsubsection{Isolated neutron stars.}
For an isolated star, apart from the timelike Killing vector $T^a$, a further assumption is the 
existence of a spatial Killing
vector  ($\phi^a$) corresponding to an azimuthal symmetry. The four-velocity of the fluid is expressed as
\begin{equation}
u^a = u^t T^a + u^t \Omega \phi^a \, ;
\end{equation}
with $\Omega$ the angular velocity of the matter as measured at infinity.
For a perfect fluid, equation (\ref{fluid1}) can be expressed in differential form as
\begin{equation}
dp - (\rho + p) (d \ln u^t - u^t u_{\phi} d \Omega) = 0 \label{bernoulli} \, ;
\end{equation}
which is referred to as the {\it relativistic Bernoulli equation}. Two cases are distinguished:
{\it uniform rotation}, $d \Omega = 0$ where equation (\ref{bernoulli}) can be trivially integrated; and
{\it differential rotation}, where the integrability condition $u^t u_{\phi}=F(\Omega)$ is used to
perform the integration. $F(\Omega)$ describes the {\it rotation law} of the matter\cite{butterworthDIFFERENTIALNS}.

The simplest model for stars were introduced by Oppenheimer and Volkoff\cite{oppie}, corresponding to non-rotating
spherically symmetric configurations parametrized by a single variable determining how relativistic the system is.
Due to Birkhoff's theorem, the solution outside the star is the Schwarzschild one. This model constitutes a valuable
test for general relativistic hydrodynamic implementations and is customarily used for this effect.

In general, isolated neutron stars will be rotating and the hydrostatic equilibrium equations must be solved
in conjunction with the constraints (\ref{eq:Hamiltonian-const},\ref{eq:momentum-const}). For uniformly
 rotating stars, the obtained solutions (for a given
equation of state) are parametrized by $\Omega$ and the value of the central density which serves as an indication
of how relativistic the solutions are. For differentially rotating stars, the rotation law must be specified.
As mentioned, data must be specified to solve the constraints and different choices have led to a number of
approaches. Some examples of them 
are\cite{wilsonISOLATEDNS,butterISOLATEDNS,friedmanISOLATEDNS,bzzolaISOLATEDNS,cookISOLATEDNS,komatsuISOLATEDNS}. 
(For a recent review on the subject see\cite{stergioulasreview}.)

\subsubsection{Binary neutron stars.}
Binary systems can not rigorously be in equilibrium as they emit gravitational radiation. However,
when the members of the binary are far apart (beyond the {\it inner most stable
circular orbit}), the gravitational radiation reaction time scale is much longer than the orbital
period and a reasonable assumption is to consider the stars are in a quasiequilibrium state. 
This state is reflected in an approximate killing vector in a frame co-rotating with the
binary. Ie. if the binary rotates with angular velocity $\Omega$, this killing vector is
\begin{equation}
\hat T^a = T^a + \Omega \xi^a \, ,
\end{equation}
where $\xi^a$ is the generator of rotations about the rotation axis and $T^a=(\partial_t)^a$. 
Numerical implementations of binary systems were initiated by Wilson and Mathews\cite{wilsonmathews}
where the fluid variables are not prescribed enforcing hydrostatic equilibrium. Rather,
an initial guess for the density profile is specified and the system is evolved until
equilibrium is reached. In order to have a clearer physical picture of the initial 
configuration hydrostatic equilibrium can be enforced at the initial time.
Work on obtaining equilibrium
configurations has concentrated on two different assumptions leading to
considerably different solutions: (I) {\it co-rotation} where $u^a \propto \hat T^a$ and the individual
stars in the binary do not rotate
with respect to the co-rotating frame defined by $\hat T^a$
and (II) {\it counter-rotation} where the individual stars do not rotate with respect to the rest frame of the binary.\\

{\bf Corrotating binaries}\\
With respect to the co-rotating frame, the stars appear to be in a (extremely slow) head-on trajectory; hydrostatic
equilibrium is specified by solving the relativistic Bernoulli equation (under the assumption $d \Omega =0$)
together with the constraints\cite{baumgarteQE}.
The main drawback of this approach has to do with its relevance for astrophysical purposes. The viscosity of the 
fluid composing the neutron stars is not expected
to be large enough viscosity for the spin to ``lock'' with the orbit (as is the case in the
earth-moon system)\cite{bildsenNOCOROTATION,kochanekNOCOROTATION}. If the spins of the
neutron stars are small, for close binaries, {\it irrotational} fluid models are expected to provide a
more reasonable approximation.\\

{\bf Irrotational binaries}\\
Irrotating (also referred to as counter-rotating) binaries are obtained assuming the matter has irrotational
 flow\cite{bonazzolaQE,teukolskyIRROT,shibataIRROT}.
This assumption allows expressing the velocity of the fluid in terms of a ``vector potential'' $\Phi$,
\begin{equation}
h u_a = \nabla_a \Phi \, ;
\end{equation}
with $h$ the enthalpy. When expressing $u_a$ this way Euler's equation (\ref{fluid1}) is automatically
satisfied, leaving only the continuity equation to be solved (\ref{fluid2}) which can be expressed
as a Poisson equation for $\Phi$. The quasistationarity condition is expressed as,
\begin{equation}
h u_a \hat T^a = constant \, ;
\end{equation}
which is readily obtained from the Killing equation\cite{teukolskyIRROT}. The continuity equation coupled with
appropriate boundary conditions at the surface of the stars and the constraints are then solved simultaneously to
yield quasiequilibrium counter-rotating configurations. Numerical implementations have been presented
in\cite{bonazzolaQE,marronettiQE,uryuQE}. \\

\subsection{Black hole/neutron star binary:}
The first (and as far as I know only one) data set describing a system containing a
non-spinning black hole and a polytrope star (which is taken to approximate the neutron star)
has been recently presented by Miller\cite{millerBHNS}.
The method combines the puncture method\cite{puncturesID} to specify the black hole with the 
assumption of corrotation to treat the fluid describing the star\cite{baumgarteQE}. It produces
accurate initial data to study the system approximatively assuming quasiequilibrium\cite{millerBHNS},
or as initial data for a complete description of the system through a 3+1 code. This is an important
first step, and will likely lead to more realistic initial data when the irrotational case is considered.\\

\section{Main accomplishments}
Perhaps the most spectacular accomplishment to date is the discovery of critical 
phenomena in General Relativity by Choptuik\cite{mattFIRSTCRITICAL} and analogous behavior 
in a wealth of different systems discovered though numerical models\cite{carstenreview}. 
This and several other important achievements illustrate the potential of Numerical Relativity; to name a few,
\begin{itemize}
\item{{\it Bagels might form when black holes collide/form}: In the early 90's Shapiro and Teukolsky studied
a system containing a toroidal distribution of particles\cite{shapiroteukolskyTORUS}. These simulations
followed the collapse
of these particles and the resulting event horizon was obtained by tracing (past directed) null rays
from the end of the simulation\cite{shapiroteukolskyTORUSEH}. Strikingly, what they found was that early phases of the horizon
topology corresponded to a {\it toroidal horizon} while at late times, as expected, to a {\it spherical horizon}.
This at first sight was puzzling as this toroidal horizon appeared to leave room for violations of
cosmic censorship. Shortly after these results, an analytical
model studying the caustic/crossover structure of null surfaces showed that indeed this toroidal topology
was the correct picture\cite{jeffshapiroteukolsky}. Cosmic censorship is not violated
as the `hole of the torus' pinches off faster than the speed of light. Additionally,
 recent analytical models
have shown that a toroidal structure of the early phase of colliding black holes might indeed be the {\it generic} 
behavior\cite{fissionwh,husaTORUS,shiinoTORUS}. 
It will be a `nice' challenge for numerical simulations to reproduce this expected feature. }
\item{{\it Head-on collision of black holes}: A two dimensional code was used to simulate the head on
collision of non-spinning black holes\cite{ncsa_2Dcomp,seidelHEADON}. Not only were these simulations 
capable of accurately follow
the evolution past merger for a decent amount of time but of extracting the gravitational waves, observe
the ring-down of the merger hole for several periods and reconstruct the event horizon
structure (revealing the expected `pair of pants' \cite{seidelEVENTHORIZONS}). These simulations were
carried out with the use of singularity avoiding slicings (Maximal slices). Additionally the obtained
results were successfully corroborated with those obtained from perturbative studies\footnote{For a review on the
subject see\cite{pullinCLAPREVIEW}}. A remarkable agreement of results obtained with both approaches 
was achieved\cite{CLAP_NUM}. These results have a twofold message, on one
hand, perturbation analysis (used in a regime where one expects it to be valid) can be used to
check a numerical implementation; on the other hand, the numerical implementation might show that the
regime of validity of the perturbative approach be larger than first expected. Obtaining ``error bars'' for
perturbative treatments is an involved process requiring working out the following order in the perturbative
expansion\cite{nicasio2nd}. A carefully tested simulation can certainly provide these error bars in a much more direct
way and 
be used to decide whether the, cheaper, perturbative method can be used to describe the system at certain stages 
.} 
\item{{\it Generic single black hole simulations}: Simulating stably a single black hole in 3D for unlimited
periods was proven possible\cite{movebh,bhforever}. Initial data corresponding to single Schwarzschild or 
Kerr black holes plus some amount of gravitational radiation was accurately simulated
for tens of thousands of $M$
($M$ being the mass of the black hole) without signs of instabilities. This work employed singularity
excision highlighting its usefulness. As a test of causality not being violated, different excision regions
were defined by choosing the apparent horizon or different types of surfaces (lying inside the apparent horizon
but not coinciding with it), physical ``measurements'' were carried out in the exterior and the solutions were checked to agree quite well.}
\item{{\it Qualitative studies of Binary Neutron Star Spacetimes:} An approach that has been exploited
to gain insight into the behavior of binary neutron star systems assumes the system is in {\it quasi-equilibrium}.
Under this approach, the system is assumed to radiate negligible amounts of energy and the
system can be, in some sense, approximated by obtaining equilibrium configurations at different
 separations\cite{bonazzolaQE,marronettiQE,baumgarteQE,uryuQE}.
This translates into solely having to solve the initial value problem (ie. find data satisfying the constraints).
This approach has been used to obtain estimates of the location of the innermost stable circular orbit (ISCO) 
and the behavior of the central densities of the stars as they approach each other, even closer than the ISCO.
It is unclear to me that this approach can be pushed {\it this far}, as at the ISCO neglecting gravitational
radiation is not consistent and its accounting by means of the quadrupole approximation might not be
accurate enough. The results predicted from this approach will eventually be corroborated
or not by fully dynamical evolutions.}
\item{{\it Singularity studies:} Understanding whether singularities are hidden, which types they
are, etc. has been
another goal of numerical investigations and important results have been obtained. \\
{\it Singularities in collapse situations:} Naked singularities in gravitational collapse of a scalar field
have been found by Choptuik\cite{mattFIRSTCRITICAL} and many others 
(see for instance\cite{hamadeCRITICAL,garfinkleCFINCHARACT}) additionally revealing a self-similar
or discrete self similar behavior of the solution\cite{carstenreview}. \\
{\it Nature of singularities in charged/rotating black holes:} Spacetimes containing
rotating or charged spacetimes possess a Cauchy horizon (CH)\cite{wald}. Studies on the effect of perturbations on this CH
were initiated (analytically) by Poisson and Israel to check conjectures
that these perturbations would drive the CH into a true singularity\cite{poissonCH}. During the last decade a number of
numerical investigations were capable of showing this is indeed the case\cite{gnedinCH,bradyCH,burkoCH,piranCH}. 
Moreover, numerical investigations provided the complete picture\cite{piranCH}; that is, generically the CH becomes
a null, weak singularity which is a precursor of a strong spacelike singularity.\\
{\it Singularities in Cosmological Models:} In homogeneous cosmologies the generic singularity is approached
either by the Kasner solution\cite{kassner} or by displaying Mixmaster dynamics\cite{beverlyREVIEW}.
Furthermore, it has been  conjectured that singularities in 
generic four dimensional space-times are spacelike and oscillatory
(Belinski, Khalatnikov and Lifschitz\cite{bkl}) while
generic space-times  with stiff fluids (including massless scalar fields) have
singularities which are spacelike and non-oscillatory (as conjectured by
Belinski and Khalatnikov\cite{bk}). Additionally, according to this picture,  
spatial points decouple near the singularity and the local behavior is asymptotically
like spatially homogeneous (Bianchi) models. Spacetimes with non-stiff matter
appear, close to the singularity, to behave independent of the matter and the evolution is
determined by the curvature. On the other hand, for stiff matter, this dominates the evolution
and is responsible for the oscillatory behavior. Valuable insight has been provided by
numerical simulations that there exists important situations where classes of spacetimes
exhibit non-oscillatory behavior at the singularity even without the presence of stiff matter.
For instance, in the Gowdy class of spacetimes, simulations showed no oscillations\cite{bergerBKL};
this result was later analytically proven\cite{rendallBKL,bergerBKL}. Aside from confirmation
or not of these conjectures (often referred to as BKL conjecture) for specific cases, numerical explorations of
cosmological singularities has provided evidence that each spatial points does evolve towards
the singularity independently\cite{beverlyREVIEW}.
} 
\item{{\it Critical Phenomena:} Ever since the discovery of critical phenomena 
by Choptuik\cite{mattFIRSTCRITICAL}, analogous
phenomena have been discovered basically in every possible imaginable (and workable) scenario and well beyond
a hundred papers on this topic have been published\footnote{For 
an up to date complete review on the subject refer to\cite{gundlachreview2,carstenreview}.}. 
Critical phenomena has been `observed in the numerical laboratory' in 
systems containing massive and massless Klein Gordon fields,
in Yang Mills theory, in spacetimes with perfect fluids, in gravitational collapse in 
Anti de Sitter spacetimes,  self gravitating non-linear sigma models,
in $6D$ (assuming spherical symmetry); in full $2D$ gravitational collapse, etc.
I can not cover here the rich aspects of this problem and I refer the reader to the latest
(and continuously updated) review in\cite{carstenreview}. Just to show the tip of the iceberg, 
I will here mention that the work presented in\cite{mattFIRSTCRITICAL} 
carefully studied the (spherically symmetric) Einstein Klein Gordon system in the verge of black hole formation.
Namely in a collapse situation, {\it two} could be the final states. Either a  black hole forms
or the field disperses away. At the boundary between black hole or star formation and dispersion a rich
phenomena was discovered, where the mass $M$ of the final collapsed black hole obeys a (by now famous)
scaling relation $M=C(p-p_*)^{\gamma}$ where $\gamma$ results completely independent of the initial data.
Moreover, the solution that gives rise to such a relation, displays a scale-periodic dependence for $p\approx p_*$.
The existence of such a phenomena was first discovered numerically and it marked the beginning of a new
research branch in numerical and analytical G.R. Most of the simulations displaying critical phenomena have
been carried out in $1D$ situations; I am ware of just two published studies displaying this phenomena
in $2D$\cite{abrahamsCFIN2D,mattADSCOLLAPS}. As a last point, it is worth remarking that these phenomena have been 
simulated with the three formulations
presented in section \ref{formalisms}. For examples of critical phenomena studied with
the `3+1'; characteristic and conformal approaches see \cite{mattFIRSTCRITICAL,mattCFINFLUIDS};
\cite{garfinkleCFINCHARACT,charac_crit} and \cite{peterFIRST}. }
\item{{\it Rapidly rotating neutron stars. Secular Instability:} Studies of rapidly
 rotating neutron stars provide
valuable information on the equation of state of matter at extremely
high densities and insight on them being sources of detectable
gravitational waves. In particular, oscillations can become unstable producing gravitational
waves that could be detectable, carrying information on the equation of state.
Uniformly rotating, incompressible stars are secularly unstable to bar mode formation; this
instability grows in the presence of some dissipative mechanism like viscosity or gravitational radiation. 
The instability
appears for critical values of $\beta$ [{\it =(rotational kinetic energy)/(gravitational binding energy)}].
This value depends on the compaction of the star, the rotation law and the dissipative mechanism. 
Instabilities driven by gravitational radiation have a critical 
value of $\beta \leq 0.14$ as observed in simulations~\cite{stergioulasSECULAR,imamuraSECULAR}. Viscosity, on the 
other hand drives
the critical $\beta$ to larger values\cite{bonazzolaSECULAR,shapiroSECULAR}.(For a detailed presentation 
of the subject see\cite{stergioulasreview}).}
\end{itemize}

\section{{\it Current main focus and results}}
Most present efforts are concentrated towards obtaining robust implementations of Einstein
equations in 3D while at the same time extracting physically relevant information with the
current (and constantly revised and improved) codes. There already exists robust 3D implementations
in the characteristic formulation but as mentioned they can not be applied to generic situations.
The main targets within this formulation are BH-NS systems and the post-merger phase of BH-BH systems.
3+1 and Conformal field equations implementations are not yet robust. Existing codes in these approaches
can evolve single black hole systems for at most $1000M$. If richer spacetimes (binary black holes, 
non-vacuum black hole spacetimes, etc) can be modeled for about the same time, useful physical information
can be extracted. Thus the current focus it not only to extend the simulation lengths (by re-examining analytical
and numerical issues) but also to apply the existing knowledge to investigate physically relevant
systems. 

Of the systems being considered, some have the additional incentive of being important for gravitational 
wave detection but certainly all  entice us
for their potential to shed light in our understanding of General Relativity in strong field scenarios and/or
global structure of spacetimes. Some of the current main projects are,
\begin{itemize}
\item{{\bf Black hole and or neutron stars simulations:} \\
Several efforts worldwide are being directed towards modeling systems containing 
black hole and/or neutron stars. These simulations will play an important role in the
detection and analysis of gravitational waves to be measured by LIGO\cite{LIGO}, VIRGO\cite{VIRGO}, 
GEO600\cite{GEO}, TAMA\cite{TAMA}, etc.
Considerable progress has been achieved in both fronts recently as the first simulations
of binary black holes\cite{golmCOLL,usCOLL} and binary neutron star systems\cite{suenNSCOLL,nakamNSCOLL}
are starting to appear. The simulations have been conceived more as a proof of concept than 
actual models of realistic scenarios. Nevertheless, they are not only useful to understand
the problems being faced by 3D numerical relativity but also are starting to give actual physical 
information.

{\it Binary Black hole simulations}\\
The first medium-lived simulations of binary black holes
were presented in\cite{seidelreview,golmCOLL}. This simulation used maximal slicing conditions
and zero shift. The (spinning) holes had masses $m$ and $M=1.5m$ (for a total $M_{ADM}=3.1$) , located
at $\pm M$ on the $y$ axis (ie. fairly close to each other) and their linear momentum 
was chosen perpendicular to the line of separation. The
runs proceeded nicely for about $30 M_{ADM}$ and the first period of the gravitational waves produced by the system
were obtained. The simulations were obtained using the BSSN approach\cite{shapiroEVOL,golmEVOL} and
outgoing boundary conditions were prescribed.
However, the initial data used\cite{puncturesID} assume conformal flatness which, as mentioned is not
well suited to astrophysically relevant cases. Additionally,
the use of maximal slicings prevents long term simulations. Current work is focused to incorporate
singularity excision techniques to extend these runs\cite{golmEXCISION}.

The other set of simulations has presented the first binary black hole simulation
with the use of singularity excision\cite{usCOLL}. Initial data corresponded to a grazing collision of (two spinning or not) equal
mass ($m$) black holes separated by $\approx 10m$ and with impact parameter of $m$.
Outer boundaries where placed at $20m$ 
from the `grid' origin and data was specified there by the `simplistic' approach. Singularities were
excised from the computational domain and the simulations run for about $15 M_{ADM}$.  It was noted
however, that as boundaries were pushed farther, longer simulations were obtained (indicating a strong
boundary influence). Initial data
was {\it not conformally flat}\cite{matznerID}. Present efforts are focused
in removing the instabilities and improving the outer boundary treatment.\\

The main messages from these preliminary simulations are: 
(I) considerable gravitational radiation might be expected from binary black hole
 simulations $\approx 1\% - 3\%$  (estimates obtained by analyzing the area of the 
 apparent horizons\cite{usCOLL,golmCOLL} and waveform extraction\cite{golmCOLL});
(II) excision techniques have shown to be capable of dealing with singularities, starting on a slice with
two separated black holes and following it well past the merger\cite{usCOLL}.\\

{\it Binary Neutron Star Simulations}\\
Models of binary neutron stars systems are also starting to produce simulations describing two
`neutron' stars to the point where the stars begin to
merge\cite{nakamuraCOLLNS,suenNSCOLL,shibataNSFULLGR}. The stars are represented by polytropes,
have equal masses and the codes have been constructed
using the `3+1' approach presented in\cite{shibnakam,baumgarteBOUND}.

In\cite{nakamuraCOLLNS}, `conformal' slicing and the pseudo-minimal distortion are 
used to prescribe the shift.
The stars have mass $M_{\odot}$, radius $6 M_{\odot}$, are initially 
separated by $24 M_{\odot}$ and initial data for co-rotating
or irrotating stars are simulated. Instabilities, apparently caused 
by the slicing condition used, terminate the runs obtained with this code when the stars are about to merge.
This simulation was extremely coarse ($\Delta x^i = M_{\odot}$) and boundaries where placed $95 M_{\odot}$ from the
center of mass. The authors are working on incorporating maximal slicing to their code and will run their
new simulations in a more powerful machine.
In\cite{suenNSCOLL}, maximal slicing is used to foliate the spacetime, the modeled stars
had mass $1.4 M_{\odot}$, radius $9 M_{\odot}$ and separated by $35 M_{\odot}$; they employed their code to investigate
a conjecture by Shapiro\cite{shapiroCONJ} about the non-occurrence of prompt collapse of head-on
collision of polytropes. The results in\cite{suenNSCOLL} display the formation of a black hole
in prompt timescales although further resolved simulations will be required to put the conclusions on
firmer grounds. The
 simulations presented in\cite{shibataNSFULLGR} describe co-rotating
equal mass polytropes in contact and were capable of describing the system for
a couple of dynamical timescales.\\

{\it Black hole-neutron star simulations}\\
An implementation targeting a binary system containing a black hole and a neutron star
is being developed with the characteristic formulation (exploiting the robustness displayed
in single black hole spacetimes)\cite{nigelNEW}. Because of the possible formation of caustics the range
of parameters (mass/radius of the star and proximity to the black hole) that can be simulated
with this approach is restricted. However, there is an interesting `window' of allowed values which would
enable studying astrophysically relevant systems and provide not only gravitational wave information
but also enable a global description of the system; investigate consequences of different equation of
state; influence of orbit precession on the produced gravitational wave; etc.\\

{\it Accretion of matter by a black hole}\\
Simulating the process of black hole accretion requires incorporating, among other things,
the dynamics of the fluid that describes the accreted material and electromagnetic fields.
Numerical models are yet to be completed to incorporate these ingredients into a fully G.R.
code. Achieving such a simulation will be expedited by the considerable experience 
gained through the use of
pseudo-Newtonian models where the gravitational effects of the
black hole are included
by modifying the gravitational potential and adopting suitable boundary
conditions\cite{ruffertACCRETION,armitageACCRETION,hawleyACCRETION}. First steps towards a fully
relativistic simulation of accretion processes are being carried out
by Papadopoulos and Font\cite{philipACCRETION}. Their model at present does not incorporate magneto-hydrodynamics
effects
but is already producing predictions which could bear observational importance. Namely, they find that
if mass accretion significantly increases the mass of the black hole during the emission of
gravitational waves, the expected damped-oscillatory radiative decay\cite{nollertQNM,kokkotasQNM}
is modulated by the mass accretion rate. This effect could be exploited by gravitational wave astronomy
to obtain valuable information on our understanding of black hole birth.\\

{\it Single black simulations:} Unfortunately, there still does not exist a code in
the 3+1 formulation capable of dealing with single black hole spacetimes for unlimited times. However,
considerable progress has been achieved in simulating such systems in 3D. Recently a number of 
efforts have extended the total simulation length to beyond $600M$\cite{generalEC,alcubierreSINGLEBH}. 
Given that the quasinormal period of gravitational waves is of order $~20M$, accurate simulations
for at least an order of magnitude longer provide quite a decent setting to study a variety of
interesting scenarios. In \cite{alcubierreSINGLEBH}, for instance, 
the study of collapse of gravitational waves onto a black hole is carried out and the produced waveforms
obtained. The evolution of the system is
obtained from the early dynamical phase to late times where the black hole has clearly settled 
into a stationary regime.\\

{\it Rapidly Rotating Neutron Star Simulations. Dynamical Instability:} Studies of 
the {\it dynamical} instability to bar-mode formation
of rapidly rotating neutron stars in full 3D are under way\cite{shibataNSFULLGR,shibataROTATINGNS}. As opposed
to the secular instability, the dynamical one is independent of dissipative mechanisms. Preliminary
simulations show the onset of instability for $\beta \sim 0.24$; which is slightly smaller than predictions
obtained from Newtonian implementations (see for instance\cite{newBAR,picketBAR}).
Estimates of the gravitational  wave amplitude and frequency are $h \sim  10^{-22}$ and $\sim 1kHz$ respectively.
Although more detailed simulations need be carried out, these results do show that fully relativistic
simulations of these systems are possible and might be valuable for gravitational wave detection.
}
\item{{\bf Dynamical GR - quasiequilibrium NS:} \\
As mentioned when discussing the quasi-stationary approximations of binary neutron star
systems one shortcoming of this approach is that the dynamics of the spacetime was neglected.
A more reliable description of this system (yet still short from the full numerical modeling
of neutron stars) has been recently proposed\cite{baumgarteNSQE} which employs the quasi-equilibrium
sequences described earlier to obtain a description of the stress energy tensor describing the stars
and `feeds it' to a full G.R. code. This approach, called `matter without matter'\cite{matterwithoutmatter}
 does, a priori, a
better job to describe the spacetime since gravitational radiation is not neglected (although its
back reaction on the sources is). However, when obtaining the equilibrium sequences a working assumption
has been that the three metric is conformally flat throughout all the sequence. In the G.R. part of
the approach (where Einstein equations are fully evolved) this is only enforced at an initial slice. It is
not clear whether this assumption holds during the evolution. Although conformal flatness is not required,
when producing the quasiequilibrium sequence a (by hand) prescription for the metric is assumed. 
Throughout the evolution, however, the dynamically evolved metric might not satisfy
this assumption. This can be easily monitored and as long as the agreement is acceptable this method
can be used to obtain a ``cheaper'' simulation.

Clearly, this will not be generically the case; nevertheless, this approach appears as
a natural step towards investigating the system in a more complete way than when using quasi-equilibrium
sequences and can serve as additional checks for the fully dynamical codes mentioned in the previous item.
}

\item{{\bf Critical Phenomena in higher dimensions:} \\
As mentioned, most of the simulations displaying critical phenomena have
been carried out in $1D$ situations. The first simulation displaying this phenomena in 2D was
presented by Abrahams and Evans\cite{abrahamsCFIN2D} shortly after Choptuik's discovery. However,
the resolution achieved was still quite low to allow for a detailed description. Recently, 2D systems have 
been revisited and preliminary results display this phenomena\cite{abrahamsCFIN2D,mattADSCOLLAPS}. However,
these simulations are still rather coarse  and have not yet the desired resolution. The use of adaptive mesh
refinement proved important in 1D, but certainly its role in higher dimensions will be crucial.}

\item{{\bf Singularity structure:}\\
General relativity clearly displays its difference with Newtonian theory in regions where
the curvature is large. In particular, in regions close to a singularity the theory displays its
full glory. What it can tell us about the structure of singularities is certainly an interesting issue.
In particular, we have seen that spacetimes in the verge of black hole formation (and therefore the
appearance of a singularity) the rich phenomenology of critical phenomena arises. We would also like
to understand the structure of singularities away from this limit case. Studying singularities via numerical
implementations is particularly difficult; in fact singularity excision/avoidance
techniques are introduced to get rid of them!
However, the promise of unraveling what Einstein's equations have to tell us in the very harshest regime is
certainly hard to resist. Answering questions about the existence of naked singularities; whether `hidden' singularities
share some properties; which character do they have (timelike, spacelike or null); etc. in generic situations
is the goal of numerical studies of spacetime singularities. These numerical simulations must be capable of
describing the singularities by the asymptotic approach to them. Describing the efforts to obtain such simulations and
what we have learned from them requires a review completely dedicated to it which goes beyond of the scope of
this review. For the interested reader I suggest starting with the comprehensive review in\cite{beverlyREVIEW}.} 

\item{{\bf Cosmology:}\\
Even though gravity is the weakest of the four fundamental forces, its long range character and
the impossibility to shield anything from its effects imply that General Relativity plays a fundamental
role governing the structure of the universe. Clearly, numerical relativity has a natural place
in efforts towards obtaining reliable models that can account for the observable universe.
These models must be capable of describing from the strong field behavior at the Big Bang epoch, 
include a possible inflation phase, accommodate for the standard model and the complex physics involved at
shortly after the Big Bang and follow the evolution
to the late time phases corresponding to clusters of galaxies formation and large scale mass fluctuations.
Cosmological simulations enjoy the benefit of comparing the obtained predictions with observations,
and will certainly play a fundamental role in our understanding of issues like the existence of
the cosmological constant; topology of the universe; initial singularity; gravitational wave interactions;
the model of structure formation; etc. 
For a recent review of computational cosmology and the role of numerical relativity refer to\cite{anninosREVIEW}.} 
\end{itemize}

\section{Working together: Complement with other approaches}
In the description of binary systems, some distinct phases can be recognized.
The first one, is an {\it adiabatic} or {\it inspiraling phase}, where the members of the binary
orbit around each other while the separation between them slowly
decreases as energy is carried away by gravitational radiation.  This phase can
be described by means of post-Newtonian\cite{POSTNEWTONIAN,buonannoPN} or
quasi-equilibrium\cite{uryuQE,baumgarteQE,bonazzolaQE,marronettiQE} methods.
This phase ends at the inner-most stable circular orbit and a second stage, known as,
{\it plunge and merger phase} takes place in which a single merged object forms 
(a black hole or a neutron star). Here, numerical simulations appear to be the only
way to obtain a complete description for generic situations. The final stage is 
the {\it ringdown phase} where the final object settles into equilibrium; perturbative
methods (around the expected equilibrium scenario) can be used to describe the
system.

Note that, since Numerical Relativity can in principle fully solve Einstein equations,
simulations could be used to model the {\it complete} problem (ie. on all three phases).
However, this is not feasible as the computational cost of such an enterprise would be tremendous.
It is preferable to have the simulations concentrate on the
plunge and merger phase and appropriately matching with the other two. Achieving this 
`transition' is not a straightforward task; several questions 
have to be addressed for such a task\\

{\bf Pre-merger.}
In the case of an inspiral phase treated with Post Newtonian approximations, the system
is described in a ``point-particle" way and the main variables are the positions, velocities
and angular momentum of these ``particles''. However, initial data for the second phase
is the geometry of an initial slice which requires a proper ``translation''. For the particular
case of non-spinning black holes, Alvi has presented\cite{alviBBHMETRIC} such a translation following
the method of\cite{blanchetTRANSLATE}. The metric presented in\cite{alviBBHMETRIC} is expressed 
in terms of a single coordinate system valid up to the apparent horizons of the black holes 
(in the co-rotating gauge suggested in\cite{bradyIBBHGAUGE}). Whether this presentation 
is well suited for a numerical implementation is not known as it has not yet been implemented.  
Such an implementation will prove very valuable as it will shed light into how the matching
strategy should proceed.

In the case of where the first stage is treated with quasi-equilibrium methods, there is no need
for such a translation since it directly provides the metric variables. Some of the
metric variables are obtained, as discussed, via a solution of the constraints while the others
are provided by hand; the main difficulty of this method is to choose these accurately. So far,
almost all methods have provided these assuming 
conformal flatness\cite{uryuQE,baumgarteQE,bonazzolaQE,marronettiQE} (the exception being\cite{uryuNONCONF},
although still restrictions on the metric are imposed). 
Information obtained from Post-Newtonian approximations should be exploited to provide more consistent data.\\

{\bf Post-merger.}
The interface with the third stage is certainly more direct as in both phases the
geometry is evolved. The difficulty lies in recognizing the background spacetime with
respect to which the perturbations are defined. For the case of black hole
spacetimes, a useful notion is that of {\it isolated horizon}\cite{ashtekarIH} which
can be used to provide  a rigorous and unique way to determine the parameters
describing the black hole. Another issue is that of gauge. Namely, the gauge employed
during the numerical simulation need not coincide with the one for the perturbative
approach. In principle, several slices of the numerical simulation can be used to
induce data on the initial hypersurface of the perturbative approach. This is not a
trivial task, and will have to be analyzed in a ``by-case" basis, since,
although perturbative approaches have been formulated for a few of well defined
slicing conditions, numerical implementations will use different slices depending on
the physical problem under consideration. Still, a number of scenarios will
presumably be simulated and the time spent writing this module can certainly be
worthwhile. Additionally, there is an extra ``added-bonus'' in handing the simulation
to a perturbative approach (aside from saving computational costs). The total
simulation length might be `extended' since the full numerical implementation  might
suffer from instabilities generated by boundary conditions, or late time exponential
modes.  If already a perturbative approach can be used where the quality of the
simulations at intermediate times is reasonable, the simpler perturbative approach might be
capable of producing longer total simulations. A recent work by Baker et. al.\cite{lazarus,loustolazarus}
(the `Lazarus approach')  has actually shown this can be the case. Namely, they have used an ADM full 3D
simulation (with maximal slicing) to model a binary black hole system. Initial data 
is defined with the
Misner solution\cite{misner} from a fairly close separation. Although the full 3D
simulation crashes a relatively short time after the holes have merged, the
perturbative approach is able to continue the simulation for essentially unlimited
times\cite{loustolazarus}. At least for this particular case, the combination of 
numerical relativity with a post-merger perturbative treatment, has simulated a binary
black hole plunge all they way to the final equilibrium stage. Work is underway to 
study astrophysically relevant scenarios, match to codes using black hole excision,
accommodate more generic slicing options, etc.

\section{The future role of numerical relativity}
As the field matures and enough computational resources become available, the role of
numerical simulations to understand the theory will become increasingly more important.
It is hard to imagine all branches where it will be employed, but certainly in astrophysical systems,
singularities, cosmology, global spacetime analysis and even quantum gravity.\\

In the particular case of astrophysical systems, it is worth noting that 
for decades progress towards achieving astrophysically relevant simulations have
proceeded in two fronts. One front concentrating efforts
towards accurately evolving the geometric variables (either assuming vacuum spacetimes or
treating the matter in an approximate way); the other pursuing accurate
simulations of the fluid variables (at the cost of treating problems where the geometry
was considered fixed or where dynamical effects could be taken care by pseudo-Newtonian
approaches). Recently, these fronts have started converging with renewed hopes for complete
studies of physical situations\cite{suenNSCOLL,nakamuraCOLLNS,shibataCOLLNS,shibataNSFULLGR} which will provide
further insight into these systems. Still, present simulations
do not incorporate a number of processes like neutrino transport, magneto hydrodynamics, etc.
Inclusion of these ingredients will greatly benefit from present simulations of systems obtained
with Newtonian or Pseudo-Newtonian models which have advanced the knowledge on how to
accommodate for them (see for instance
\cite{ruffertACCRETION,armitageACCRETION,hawleyACCRETION,shapiroNEWTONIAN,centrellaNEWTONIAN,rasioNEWTONIAN}). \\

{\bf Black holes; Neutron Stars and beyond} \\
Clearly, any system involving black holes or neutron stars can only be accurately studied by
taking into account General Relativity. In systems involving a single BH or NS with other
much gravitationally weaker and smaller object, the latter can be reasonably well 
represented by a point particle following a geodesic path on the spacetime defined 
by the BH or NS, see for instance \cite{hughesPARTICLE,ruoffPARTICLE} (where the backreaction of the `particle' 
is accounted for by prescriptions like those presented in\cite{mino,quin}) 
%Standard results of geodesics in Kerr or Schwarzschild spacetimes\cite{mtw,wald} can be used to 
%describe the system if the massive object is a BH. For NS case, a s
A very different treatment is needed if the system contains binaries (BH-BH; BH-NS; NS-NS) or if
the single object is surrounded by a massive accretion disk. X-ray observations already predict a
significant abundance of NS-NS and massive accretion disks and quite reasonable models predict
a considerable number of BH-NS and BH-BH binaries\cite{sciamaBH,zwartBH}. A complete study of these systems
require full 3D numerical simulations, which will not only provide important insights on their
gravitational wave output but also on the equation of state (for the NS case); active galactic
nuclei and quasars; formation of black holes; models of gamma-ray bursts (GRB's) and strong field gravity.

These simulations must incorporate general relativity, neutrino processes, magnetohydrodynamics
and nucleosynthesis and will certainly be quite a challenge for many years to come. However, their
pay-off will make the effort very much worthwhile; among them,
\begin{itemize}
\item{{\it Gravitational Waves:} Prediction of the gravitational waves from these systems will enable
deciphering the information encoded in these waves and let us understand the source system. 
Masses, spins,
equation of state, accretion rate, etc. can be readily estimated from the detected waveforms
(see for 
instance\cite{cutlerDECIPHERWAVES,vallisneriDECIPHERWAVES,finnDECIPHERWAVES,ferrariDECIPHERWAVES,kokkotasDECIPHERWAVES}).}
\item{{\it Merger recoil estimation:} In the coalescence of these strong field binaries a non-zero recoil
will result from the linear momentum carried away by gravitational waves. This effect might be particularly
relevant in the case of supermassive black holes believed to exist in most galaxies. When two galaxies collide
(and present models predict those at $z\ge3$ participated in a series of mergers!\cite{reesREVIEW}); 
the non-zero recoil velocity could be large enough that the resulting hole be dislodged from the center of
the merged galaxy. This effect would explain low-z quasars asymmetrically located in their host galaxies.
Even more spectacular, the recoil might result large enough to eject it out of the galaxy!\cite{reesREVIEW}. Only through
numerical simulations will this recoil be quantized.}
\item{{\it Black hole birth description:} Gamma Ray Bursts are for a very short time the brightest objects in the universe (much
more than the rest of the universe combined). This hints of extreme conditions causing them, and their
understanding will tell us a great deal about GR in strong field cases. One model for GRB's is that they are produced when
a massive disk ($0.1 M_{\odot}$) is accreted on to a BH\cite{piranGRB}. NS and NS-white dwarf binaries can yield precisely
these kind of situations (as could BH systems), thus GRB's might also be signaling
the birth of a black hole. Numerical simulations
of these systems will provide the ultimate corroboration of this model.}
\item{{\it Energetics of GRB's:} Although the afterglow of GRB's is well described by current models\cite{mezaros}; these
models overestimate the GRB energy\cite{piranGRB}. Understanding this issue through numerical simulations
will certainly be quite a challenge, but a computational approach might be the most reliable way to fully resolve it.}
\end{itemize}

{\bf Naked singularities}\\
Since the early attempts to produce naked spindle singularities by Shapiro and
Teukolsky\cite{shapiroteukolskyNAKEDSPINDLE} an unresolved controversy has existed.
On one hand, analytical evidence against the formation of spindle singularities
has been presented\cite{cruschielNAKEDSPINDLE}. Additionally, it has been conjectured\cite{nakamuraNAKEDSPINDLE}
that if this type of singularity could exist, it would dissapear or become a black hole by
the back reaction of the gravitational waves emitted at the formation of the singularity.
On the other hand, numerical
investigations\cite{shapiroteukolskyNAKEDSPINDLE,shapiroteukolskyNAKEDSPINDLEII,shapiroteukolskyNAKEDSPINDLEIII}
point towards their existence {\bf assuming} the failure to locate an apparent horizon
is a good indicator. However, even Schwarszchild spacetime admits slicings without apparent horizons\cite{waldNOAPHOR},
Wald\cite{waldHOOP} suggests that the singularities found in these simulations are not naked
and the apparent horizon has not yet appeared in the slicings considered. This tension can be resolved
by further numerical studies, under different slicing conditions and by analyzing the
structure of the event horizons. \\

{\bf Quantum Gravity?}\\
Numerical relativity is making its first steps into the realm of String Theory. Computational
investigations of the AdS/CFT duality are under way to analyze low enegy gravity processes
and their relation to high energy phenomena in Yang-Mills theory. A thorough understanding
of such situations would hopefully contribute to the understanding 
of the subject\cite{husainCHARACT}. Additionally
simulations about stability of black strings are also being considered. 
As first noted
by Gregory and Laflamme a notable difference of gravity in higher dimensions is that
black holes are not stable\cite{gregorylaflamme}. By perturbative calculations, these authors showed that a `black string'
(the higher dimensional analogue of a black hole) 
is not stable under perturbations. Due to their analysis being restricted to 
linearized perturbations, it was not clear what the `final' fate of these perturbed black string was.
Recently, in the case were certain assumptions are satisfied, it has recently been shown that
the horizon does not pinch off
but rather it apparently settles into some new static black string solution\cite{horowitz}.
On a separate treatment, (one which does not require the assumptions in\cite{horowitz} be satisfied),
it has been argued through a linear perturbation analysis and and a Newtonian analysis that the final
fate corresponds to a collapse of the spacetime in the string direction\cite{unruhstring}.
A full numerical solution would certainly shed light on this problem by revealing this 
final state. Preliminary  studies of this problem, under the assumption of spherical symmetry,
are being carried out with a 2+1 code (ie. radius and `string' coordinate + time) 
and hopefully will report interesting results in the near future.\\

{\bf ``Conjecture-testing''}\\
Physical intuition has given rise to a number of conjectures, among them: Cosmological
Censorship\cite{hawkell}; `Hoop' conjecture\cite{hoop}; Belinski-Khalatnikov-Lifschitz conjecture\cite{bkl,bk};
`Shapiro conjecture'\cite{shapiroCONJECTURE}; etc. which have proven very difficult to
prove (or disprove). Numerical simulations can shed light on their
validity; in particular, they have already shown the possible 
existence of naked 
singularities\cite{mattFIRSTCRITICAL} and Cauchy horizons being driven to true 
singularities\cite{piranCH}. \\

{\bf Global spacetime structure} \\
Numerical relativity can play an important role in global properties of spacetimes with
isolated sources. Penrose's realization of asymptotically simple (AS) spacetimes, shows the  
relationship between Einstein's equations, geometric asymptotics, conformal geometry and
the notion of isolated system\cite{penrose}. In particular, the concept of asymptotically simplicity implies the 
Weyl tensor displays a {\it ``peeling behavior''} of the Weyl curvature and since its 
introduction a recurrent issue in General Relativity has been how general it is.
A well known system displaying a peeling behavior weaker than that implied by an AS spacetime is
the `perturbed' Minkowski spacetime studied by Christodoulou and Klainerman\cite{chrisklain}
(perhaps a restriction on the initial data considered in\cite{chrisklain} might yield an AS spacetime).
Numerical investigations might provide valuable indications on spacetime properties on the large;
a rigorous analysis would demand being able to simulate the whole spacetime; at present,
it appears the conformal field (section \ref{form:conformal}) and Cauchy-characteristic
matching approaches are the best suited for such a task.
Less ambitious estimates, but likely useful ones, can still be made with implementations
in the 3+1 formulations by studying the fields in the far zone.
\\

\section{Conclusions}
In the present review lack of space has prevented me from addressing
every subject in detail; thus, I have intended this work to be an up to date `tour' 
through the many aspects present in today's Numerical Relativity research. In some cases, 
I have chosen to briefly describe
the goal and main aspects of: Relativistic Hydrodynamics\cite{fontreview}; Computational 
Cosmology\cite{anninosREVIEW}; Singularity Studies\cite{beverlyREVIEW} 
and Critical Phenomena\cite{gundlachreview2} and refer the reader to recent reviews on
these subjects.

I have discussed the several `flavors' presently found in numerical relativity.
Both from their approach towards Einstein equations
and their numerical strategies to implement them. I have tried to emphasize
the ideas, techniques and main problems together with the main accomplishments
and outstanding problems which will keep everyone quite busy in the coming years.
Yet, this list is by no means exhaustive, we still do not know what treasures have
been kept hidden in the theory waiting for us to discover. Certainly, the road in front of
us is not an easy one, but is likely to be one with exciting discoveries. As the
(translated) words of Antonio Machado tell us: ``Traveller there are no paths, paths are made
by walking''.

\ack{
Special thanks to M. Alcubierre, N. Bishop, M. Choptuik, H. Friedrich, C. Lousto, 
P. Marronetti, R. Matzner, J. Pullin, S. Husa and J. Winicour
for helpful comments
and a careful reading of early versions of the manuscript.
I have enjoyed discussing points of this review with members of the Numerical Relativity
groups at the University of British Columbia and The University of Texas at Austin.
I would like to express my gratitude to many who have informed me of their latest efforts: K. Alvi, 
T. Baumgarte, B. Bruegmann, J. Bardeen, D. Choi, J. Frauendiener, C. Gundlach, R. D'Inverno, 
P. Diener, D. Garfinkle,
P. Huebner, P. Laguna, M. Miller, P. Papadopoulos, M.
Scheel, W.M. Suen., D. Shoemaker, J. Thornburg and M. Tiglio.}

%\newpage
%\thispagestyle{empty}
%\vspace*{279pt}
%\begin{center}
%                {\Large \bf BIBLIOGRAPHY}
%\end{center}

%\newpage
\section*{References}
%\bibliography{refer_review}
%\bibliographystyle{acm}
%\bibliographystyle{trynew}

\end{document}